\documentclass[lettersize,journal]{IEEEtran}
\usepackage{amsmath,amsfonts}
\usepackage{algorithmic}
\usepackage{algorithm}
\usepackage{array}
\usepackage[caption=false,font=normalsize,labelfont=sf,textfont=sf]{subfig}
\usepackage{textcomp}
\usepackage{stfloats}
\usepackage{url}
\usepackage{verbatim}
\usepackage{graphicx}
\usepackage{cite}
\usepackage{xcolor}

\newcommand{\add}[1] {\textcolor{black}{#1}} 
\newcommand{\blue}[1] {\textcolor{blue}{#1}} 
\usepackage{multirow}
\usepackage{multicol}
\usepackage{booktabs}
\usepackage{lipsum}
\usepackage{marginnote}
\usepackage{balance}

\begin{document}

\title{Mixture of Multicenter Experts in Multimodal AI for Debiased Radiotherapy Target Delineation}
\author{Yujin Oh$^{1\ast}$, 
Sangjoon Park$^{2,3\ast}$, 
Xiang Li$^{4\ast}$, 
{Pengfei Jin$^{4}$},
Yi Wang$^{5}$, 
Jonathan Paly$^{5}$,
Jason Efstathiou$^{5}$, \\
Annie Chan$^{5}$, 
Jun Won Kim$^{6}$, 
Hwa Kyung Byun$^{7}$, 
Ik Jae Lee$^{2}$,
Jaeho Cho$^{2}$, 
Chan Woo Wee$^{2}$, \\
Peng Shu$^{8}$, 
Peilong Wang$^{9}$,
Caiwen Jiang$^{9}$,
Nathan Yu$^{9}$,
Jason Holmes$^{9}$,
Jong Chul Ye$^{10}$, \IEEEmembership{Fellow, IEEE},\\
Quanzheng Li$^{3\dagger}$, 
Wei Liu$^{9\dagger}$, 
Woong Sub Koom$^{2\dagger}$,
Jin Sung Kim$^{2\dagger}$, and
Kyungsang Kim$^{4\dagger}$

\thanks{$^{*}$: Co-first authors with equal contribution. $^{\dagger}$: Co-corresponding authors. \\
$^{1}$Department of Biomedical Systems Informatics, College of Medicine,
Yonsei University, Seoul, Korea, 
$^{2}$Department of Radiation Oncology, Yonsei University College of Medicine, Seoul, Korea, 
$^{3}$Institute for Innovation in Digital Healthcare, Yonsei University, Seoul, Korea, 
$^{4}$Center for Advanced Medical Computing and Analysis (CAMCA), Department of Radiology, Massachusetts General Hospital (MGH) and Harvard Medical School, Boston, MA, USA, 
$^{5}$Department of Radiation Oncology, Massachusetts General Hospital, Boston, MA, USA, 
$^{6}$Department of Radiation Oncology, Gangnam Severance Hospital, Seoul, Korea, 
$^{7}$Department of Radiation Oncology, Yongin Severance Hospital, Yongin, Korea, 
$^{8}$School of Computing, University of Georgia, GA, USA, 
$^{9}$Department of Radiation Oncology, Mayo Clinic, AZ, USA, 
$^{10}$School of AI, KAIST, Daejeon, Korea. Email: kkim24@mgh.harvard.edu.}
\thanks{The visual abstract of this study was supported by Research Affairs, Yonsei University College of Medicine.}
\thanks{\add{This paper extends the work (Oh, Jin and Park et al., 2025) presented at the Forty-Second International Conference on Machine Learning (ICML) 2025.}}
}

\maketitle



\maketitle

\begin{abstract}
Clinical decision-making reflects diverse strategies shaped by regional patient populations and institutional protocols. However, most existing medical artificial intelligence (AI) models are trained on highly prevalent data patterns, which reinforces biases and fails to capture the breadth of clinical expertise. Inspired by the recent advances in Mixture of Experts (MoE), we propose a Mixture of Multicenter Experts (MoME) framework to address AI bias in the medical domain without requiring data sharing across institutions. MoME integrates specialized expertise from diverse clinical strategies to enhance model generalizability and adaptability across medical centers. We validate this framework using a multimodal target volume delineation model for prostate cancer radiotherapy. With few-shot training that combines imaging and clinical notes from each center, the model outperformed baselines, particularly in settings with high inter-center variability or limited data availability. Furthermore, MoME enables model customization to local clinical preferences without cross-institutional data exchange, making it especially suitable for resource-constrained settings while promoting broadly generalizable medical AI.
\end{abstract}


\section{Introduction}
\begin{figure*}[!t]
\centering
\includegraphics[width=0.83\linewidth]{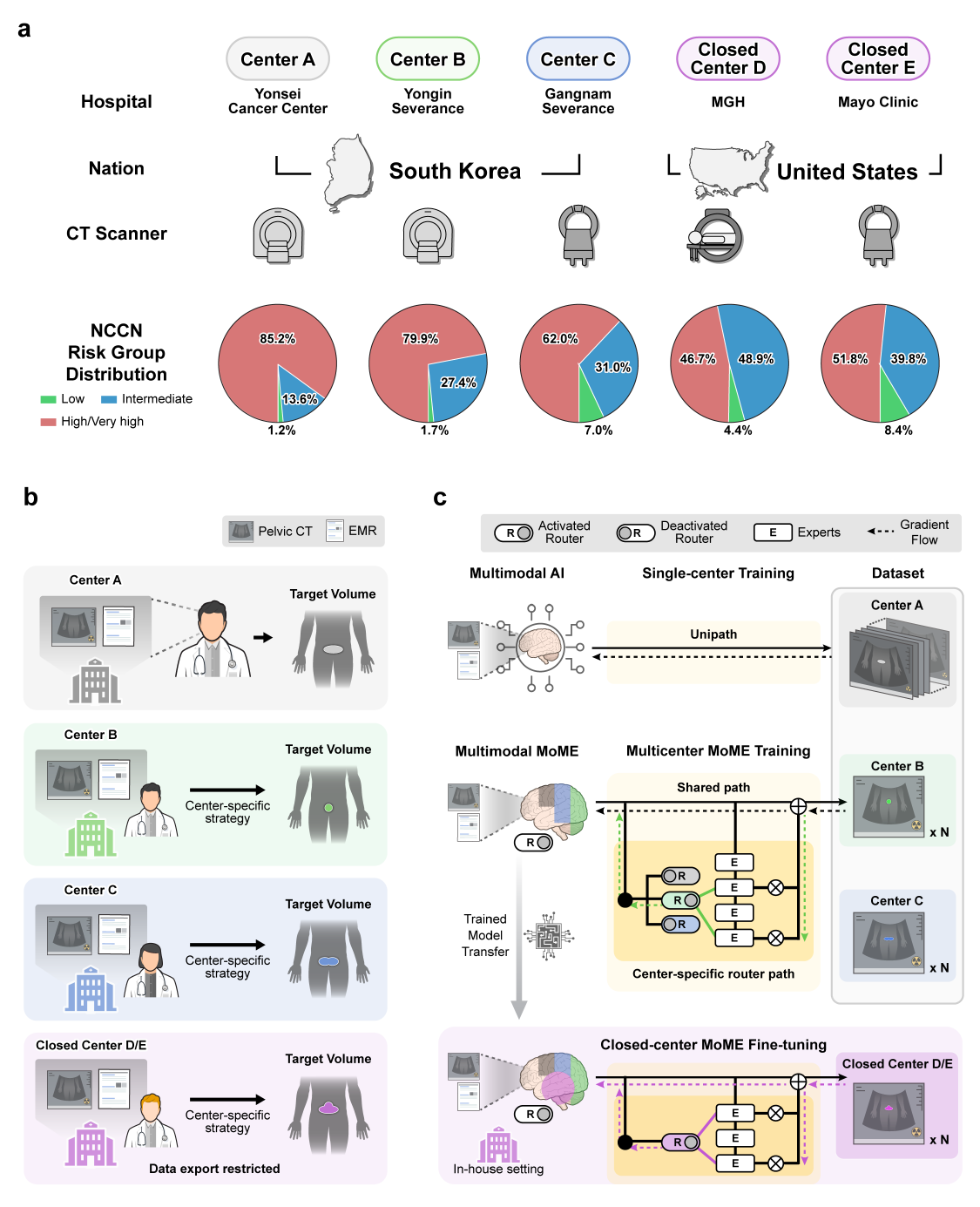}%
\caption{Schematic of multicenter AI training using our proposed Mixture of Multicenter Experts (MoME) approach.
(a) Characteristics of each center, which can influence their radiotherapy target delineation tendencies, emphasize the need for a debiased AI training approach.
(b) Radiotherapy target delineation strategies for prostate cancer patients vary across various centers, which limits the generalizability of AI models.
(c) Compared to traditional unipath single-center training, our MoME training leverages both shared and center-specific routing paths. These paths activate relevant expert modules customized to the unique characteristics of each center given a few-shot dataset. Furthermore, in hospitals where data export is restricted, a closed center MoME fine-tuning approach is employed, enabling model adaptation to the local in-house setting using only a few-shot dataset.}
\label{fig_intro}
\end{figure*}

\IEEEPARstart{T}{he} integration of artificial intelligence (AI) into clinical practice is increasingly recognized for its potential to improve patient care, particularly in fields where precision is critical, such as radiation oncology \cite{huynh2020artificial, liu2023artificial}. AI has shown promise in automating and improving critical aspects of radiation therapy, such as target volume contouring and treatment planning, including determining the scope and dose of treatment from a patient's planning computed tomography (CT) scan \cite{harrison2022machine, oh2024llm, rajendran2024auto}. However, a significant challenge remains: ensuring the generalizability of AI models in diverse institutional healthcare settings. As shown in Fig.~\ref{fig_intro}(a), variations between centers, such as differences in regional populations, imaging modalities, and clinical protocols, contribute to the difficulty of applying pre-trained AI models developed in one context to distinct data distributions in others.

Recent advancements have begun to tackle this challenge by incorporating multimodal data considerations into AI models. In radiation therapy, target volume delineation requires more than just visual cues; factors such as patient's surgical history, pathology, and disease-specific biomarker levels are also essential. Multimodal AI models, which combine clinical context with imaging data, have demonstrated superior generalization capabilities across various datasets compared to their unimodal models. This is attributed to the crucial role of clinical text, typically presented in a structured format, in improving the generalizability of AI models across various types of datasets. The promising results of multimodal models have been demonstrated in various types of cancer \cite{oh2024llm, rajendran2024auto}. Moreover, the advancement of large language models (LLMs) in medicine \cite{zhang2024generalist} has accelerated the development of multimodal AI, improving generalizability across different imaging modalities and institutional settings.

Despite these advancements, traditional AI models trained on data from a limited number of institutions continue to suffer from biases that reflect the characteristics of those specific settings. This bias hinders the adaptability of AI models to diverse clinical settings, resulting in skewed predictions and leading to suboptimal performance. Addressing this issue is especially critical, particularly in radiation therapy, where there is substantial variability in target volume delineation practices, even with consensus guidelines \cite{fotina2012critical, vinod2016review, caravatta2014inter}. Prostate cancer radiotherapy is a prime example, as treatment strategies can vary considerably across institutions, driven by differences in regional patient populations and institutional protocols \cite{barkati2016magnetic, valicenti1999variation}, as illustrated in Fig.~\ref{fig_intro}(b).  This variability complicates the implementation of AI-driven tools for target volume contouring, compared to the relatively broader acceptance of AI for contouring organs-at-risk (OAR) \cite{shi2022deep, zhang2023segment}.

In this study, we propose a Mixture of Multicenter Experts (MoME)  as a novel debiasing AI training approach to address biased inference and enable AI models to better reflect the needs of individual institutions. 
The MoME can not only mitigate data bias but also improve the generalizability and adaptability of medical AI, expanding its applicability across diverse clinical settings. 
The proposed MoME framework integrates a shared path with center-specific router paths, enabling the model to adapt to diverse data distributions and clinical settings with minimal data input. As illustrated in {Fig.~\ref{fig_intro}(c)}, this design allows the model to efficiently adapt to each medical center data distribution with using only a small dataset—ranging from 10 to 20 computed tomography (CT) scans with pre-annotated target volumes—by leveraging the router within the MoME framework. This is a significant improvement over traditional training methods, which often require hundreds or even thousands of labeled datasets to mitigate model bias to dominant institution's data distribution. This adaptability ensures that the model can account for the unique treatment approaches and delineation strategies of each institution, resulting in more personalized and precise treatment planning. 
During deployment, the MoME framework can quickly adjust to local practices and patient populations by using a few sample test datasets from a new center. Crucially, the scalability of the MoME framework is enhanced by its distributed model weights, \footnote{\url{https://github.com/tvseg/MoME-RO}}, 
facilitating integration across multiple institutions globally and allowing for the selection of the most relevant inference scenarios tailored to their practices. 

We apply the proposed MoME framework to address the limitations of institutional biases in existing AI models for prostate cancer target volume delineation.
Moreover, we extend our approach to closed center MoME fine-tuning by utilizing in-house datasets from hospitals with restricted data-sharing policies. This addresses the challenges of clinical deployment when adapting AI models to new data distributions. 
Our results demonstrate that a MoME-based model not only significantly outperforms traditional AI models in target volume contouring, but also aligns its overall distribution more closely with that of each institution. 
This improvement highlights the potential of the MoME approach to advance AI adoption by addressing debiasing challenges and adapting to subtle variations in institutional treatment protocols and patient distributions across different sites. Additionally, the modular design of MoME framework allows it to serve as a plug-in component for various AI systems, supporting continuous improvement through the seamless integration of new data from diverse sources.


\section{Related Work}

\subsection{Debiasing in Medical AI Training}
{Bias is a prevalent challenge in medical datasets, particularly due to institutional differences and patient distribution disparities, which can cause medical AI models to produce skewed predictions aligned with the distribution of their pretraining datasets. Addressing bias in medical AI requires robust training strategies. Data augmentation methods aim to mitigate bias by expanding underrepresented distributions \cite{ktena2024generative}; however, generating diverse samples from skewed distributions is computationally inefficient and challenging for high-dimensional medical image. 
Recent studies have explored loss-level fairness adjustments such as Fair Error-Bound Scaling (FEBS) \cite{tianfairseg}, and distributionally robust strategies including multi-expert \cite{jeong2026multiexpert} and group distributionally robust optimization \cite{panaganti2026groupd} approaches. However, such methods remain susceptible to data distribution manipulation and may compromise accuracy. Bias from multi-institutional differences can instead be mitigated through multicenter training that integrates diverse data sources while preserving confidentiality.
Federated learning addresses restricted clinical data sharing by decentralizing storage and enabling collaboration across institutions without directly exchanging sensitive data \cite{mcmahan2023fedavg, li2020fedprox}. However, its practical adoption remains limited: performance often falls short of centralized training, straggler problems introduce instability, and vulnerability to security threats such as data poisoning and inference attacks further constrains its use.
Recent advancements in the Mixture of Experts (MoE) training mechanism \cite{shazeer2017outrageously} have revolutionized the adaptation of AI models to diverse data distributions, particularly within continual learning frameworks. MoE significantly improves robustness and adaptability when addressing previously unseen data patterns \cite{van2020brain, rypesc2024divide, yu2024boosting}. Leveraging these innovations, we introduce the Mixture of Multicenter Experts (MoME), a novel approach designed to tackle debiasing challenges in medical AI by accommodating the variability inherent in multicenter datasets.
}

\subsection{AI for Radiotherapy Target Delineation}
{
In radiation oncology, treatment target volumes are categorized into Gross Tumor Volume (GTV), Clinical Target Volume (CTV), and Planning Target Volume (PTV). GTV represents the observable tumor, typically defined using imaging modalities and aligning closely with traditional segmentation tasks. CTV encompasses areas at risk of microscopic disease beyond the GTV, determined by tumor type, histopathological findings, TNM staging, and patient-specific factors such as age and performance status. PTV expands CTV to account for positional uncertainties during treatment \cite{burnet2004defining}.
Modern CT-based treatment planning requires meticulous delineation of target volumes and OARs across all CT slices for accurate dose calculation and planning. This process is labor-intensive, highlighting the need for AI-based solutions to enhance efficiency and precision. Early AI applications primarily focused on OARs segmentation, with deep learning models since 2016 achieving high accuracy in delineating dozens of OARs \cite{tang2019clinically}, leading to clinically impactful commercial tools.
}

{However, AI solutions for target volume delineation remain underdeveloped. Existing models are predominantly anatomy-based, targeting predefined nodal areas such as axillary, internal mammary, and supraclavicular lymph nodes in breast cancer, neck nodes in head and neck cancer, or pelvic nodes in pelvic cancers \cite{lin2021deep}. These models, while guided by standardized guidelines, often lack the integration of clinical context, limiting their applicability in patient-specific scenarios. Significant variability in clinical practice further complicates AI development, particularly for complex CTV delineation, which requires integrating disease extent and pathological findings rather than relying solely on anatomical features. Variations across institutions, countries, and individual physicians pose challenges to creating universally accepted AI models for radiotherapy. For widespread clinical adoption, AI models must adapt to diverse practice patterns and accommodate institution- and physician-specific preferences. Addressing these variations is essential to developing robust, clinically relevant AI-driven solutions for radiotherapy target volume delineation.
}

\subsection{Target volume delineation in prostate cancer radiotherapy}

Radiotherapy for prostate cancer is employed with definitive, salvage, or palliative intent. Definitive radiotherapy serves as a curative option for patients unable to undergo surgery due to advanced age, comorbidities, or personal preference. Some patients also choose radiotherapy over radical prostatectomy despite surgical feasibility. Salvage radiotherapy is used post-surgery for rising prostate-specific antigen (PSA) levels or confirmed recurrence, while palliative radiotherapy manages metastatic disease, such as bone metastases, with highly variable target delineation depending on clinical scenarios \cite{NCCN2024Prostate}.
In definitive radiotherapy, the target typically includes the prostate, seminal vesicles, and suspected extracapsular extensions \cite{salembier2018estro}. Postoperative radiotherapy targets the prostate bed and seminal vesicle bed, incorporating anatomical considerations from the surgical field \cite{dal2023estro}. If lymph node involvement is confirmed, or if the patient falls within the high-risk or very high-risk groups according to National Comprehensive Cancer Network (NCCN) guidelines \cite{NCCN2024Prostate}, pelvic nodal irradiation (PNI) is recommended, even in the absence of radiographic evidence of lymph node metastasis. Intermediate-risk patients with unfavorable factors may also receive PNI based on institutional protocols, though practices vary. Despite general principles guiding target volume delineation \cite{salembier2018estro, NCCN2024Prostate}, variability persists, particularly regarding margins and inclusion of adjacent structures in suspected locoregional invasion.

\section{Methods}

\subsection{Dataset characteristics and clinical context}
\label{met_data}


In this study, we utilize datasets from five centers located in different countries, as illustrated in Fig.~\ref{fig_intro}(a). Detailed information regarding the number of patients, tumor stage, histopathological grading, PSA levels, surgical status, treatment intent, and imaging acquisition protocols for each center is provided in Supplementary Section I and Supplementary Table I. To provide relevant clinical context, we extract key factors essential for prostate cancer radiotherapy from the electronic medical records (EMRs). These factors are chosen based on their importance for treatment planning and their availability across all institutions. The curated data are standardized into a formatted clinical dataset, as shown in Supplementary Table II.

\subsection{Multimodal MoME framework}

\begin{figure*}[!bt]
\centering
\includegraphics[width=0.9\linewidth]{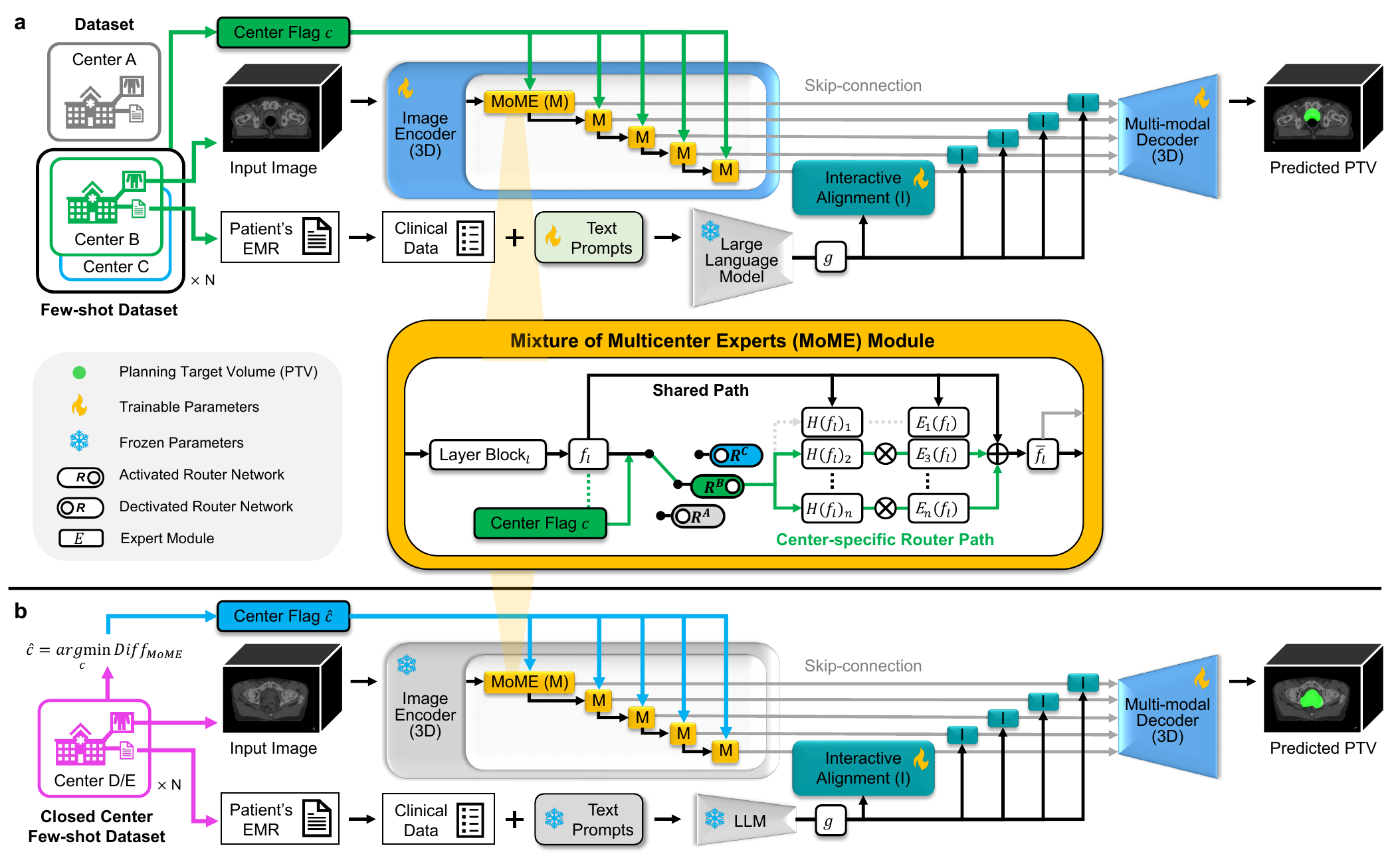}%
\caption{{Schematic of multimodal mixture of multicenter experts (MoME) framework. (a) Center-specific MoME training is performed given multiple center dataset. The center flag $c$ splits the layer-wise image embeddings $f_{l}$ into a shared path along with a center-specific router path, which are combined at the end of the MoME module to yield $\bar{f_{l}}$. For multimodal target contouring, patient's EMR is curated to yield the clinical data, followed by text prompt tuning of the frozen LLM to yield the context embedding $g$. The context embedding is then interactively aligned with the image embeddings, and decoded to yield the planning target volume (PTV) prediction. (b) Closed center MoME fine-tuning is performed while the pre-trained encoder and text prompts kept frozen. }}
\label{fig_schematic}
\end{figure*}

Our framework builds upon the sparse MoE mechanism  \cite{shazeer2017outrageously}, but unlike conventional MoE, it is tailored for multicenter debiased training by extending {our prior distribution-aware MoE (dMoE) framework \cite{oh2025dmoe}. While dMoE is unimodal and uses disease severity as a fairness factor, the proposed MoME integrates multimodal clinical data including multiple severity factors via an LLM and incorporates center information for debiased routing (Fig.~\ref{fig_schematic}). This shifts the focus from single-factor bias to center-specific distribution modeling, enabling debiased learning across heterogeneous multicenter data.}

\add{For multicenter training without full data sharing, only a small few-shot subset (upto 3-shots) from each center is shared with the main training center, while in closed center scenarios no data is exchanged and only network parameters are shared. During multicenter training, all encoder and decoder parameters are shared globally across centers, avoiding center-specific overfitting. The only center-dependent component is the MoME router, which selects top-$k$ expert modules conditioned on the center flag. This design enables shared feature learning while preserving center-specific debiasing through expert routing.}
Our multimodal MoME framework consists of \add{four key steps}: 1) center-specific MoME training, 2) fine-grained multimodal alignment, 3) center-specific MoME inference, and 4) closed center MoME fine-tuning.

\vspace{10pt}

\subsubsection{Center-specific MoME training}  

\add{Center-specific MoME training integrates multiple center-specific router networks \( R^c \) and a shared set of \( n \) expert modules, consisting of shallow multi-layer perceptron (MLP) neural networks, defined as \( E_n\).} During training, as illustrated in Fig.~\ref{fig_schematic}(a), given $l$-th layer image embeddings ${{f_l}}$ and a center flag $c  \in \{A, B, C\}$ for the corresponding datasets from Centers A, B, and C, respectively, the activated center-specific router $R^c$ selects the top-$k$ experts and computes a weighted summation of their outputs:
\begin{align}
\bar{f_l} =  {f_l} + \sum_{i=1}^{k} {R^c({f_l})_i} \cdot {E_i({f_l})},
\label{eq_exeprt}
\end{align}
where $R^c()$ outputs a weight matrix that prioritizes each expert’s contribution in a center-specific manner. The resulting weighted output is then combined with ${f_l}$, the shared path representation, to produce the final MoME image embedding $\bar{f_l} \in \mathbb{R}^{H_l W_l S_l \times Ch_l}$. The router network $R^c$ computes the sparse weight $H$ using Gaussian noise, as follows:
\begin{equation}
R^c(x) = \text{Softmax}(\text{KeepTop-$k$}(H(x), k)),
\end{equation}
\begin{equation}
H(x)_i = (x ^\top \cdot W)_i + \mathcal{N}(0, 1) \cdot \text{Softplus}((x ^\top \cdot W^{\text{noise}})_{i}),
\end{equation}
\begin{equation}
\text{KeepTop-$k$}(v, k)_i = 
\begin{cases} 
v_i & \text{if } v_i \text{ is in top } k \text{ elements of } v, \\
-\infty & \text{otherwise.}
\end{cases}
\end{equation}
where $W$ and $W^{\text{noise}}$ denote trainable weight matrices, $\text{KeepTop-}k(\cdot)$ retains the top-$k$ expert contributions, and $\text{Softmax}(\cdot)$ normalizes the selected weights.

\vspace{10pt}

\subsubsection{Fine-grained multimodal alignment}  

Following the MoME modules, the image embeddings $\bar{f_l}$ are passed to {the layer-wise interactive alignment module for multimodal alignment. We initially utilize {a local LLM to process electronic medical records (EMRs)  into structured input clinical data, as detailed in Supplementary  Table II.} To integrate clinical data during network training, we employ a second LLM. To efficiently fine-tune the LLM within our framework, we adopt text prompt tuning, leveraging learnable text prompts, extending our prior work \cite{oh2024llm}. We introduce $M$ learnable vectors, ($ \mathbf{z}_{\theta} = \{\mathbf{z}_{\theta}^1, \mathbf{z}_{\theta}^2, \ldots, \mathbf{z}_{\theta}^M\} $ parameterized by $\theta$, where $ \mathbf{z}_{\theta} \in \mathbb{R}^{M \times C} $, and $ C $ is the embedding dimension. These vectors are initialized randomly and optimized during training.
Each input clinical data $ \mathbf{s} \in \mathbb{R}^{(L-M) \times C} $ is embedded to match the dimension $ C $ of the text prompts and concatenated to form the prompted input $ \mathbf{t} \in \mathbb{R}^{L \times C} $, defined as:
\begin{align}
\mathbf{t} = \{\mathbf{z}_{\theta}^1, \mathbf{z}_{\theta}^2, \ldots, \mathbf{z}_{\theta}^M, \mathbf{s}^1, \mathbf{s}^2, \ldots, \mathbf{s}^{(L-M)}\},
\end{align}
where \( L \) is the total number of input tokens.
The prompted input \( \mathbf{t} \) is then passed through the frozen LLM, which projects it into \( L \) token-wise context embeddings \( g \in \mathbb{R}^{L \times D} \), where \( D \) is the embedding dimension of the LLM.  To align these context embeddings $g $ with the image embedding $\bar{f_l}$, we first project $g $ to match the dimensions of each $\bar{f_{l}}$ using a layer-wise linear transformation. Then, these linearly projected context embeddings ${\bar{g}}_l \in {\mathbb{R}^{L \times Ch_l}}$ are subsequently processed through self-attention and cross-attention mechanisms with ${\bar{f_l}}$ within two-way transformer modules of SAM \cite{kirillov2023segment}, resulting in multimodal image embeddings ${\tilde{f_l}} \in \mathbb{R}^{H_l W_l S_l \times Ch_l}$. These multimodal image embeddings are inputted to the decoder module, which predicts the final context-aware prediction $\hat{y}$}. The network is optimized using a combination of cross-entropy (CE) loss and the Dice coefficient (Dice) losses: 
\begin{align}
\min_{\mathcal{M}, \theta} \mathcal{L} = \lambda_{\text{ce}} \mathcal{L}_{\text{ce}}(\hat{y}, y)  + \lambda_{\text{dice}}\mathcal{L}_{\text{dice}}(\hat{y}, y) , 
\label{losses1}
\end{align}
where $\mathcal{M}$ represents our proposed multimodal MoME framework, $\theta$ denotes the learnable text prompt parameters, $y \in \mathbb{R}^{B \times H W S \times C}$ is the ground-truth PTV mask, and the predicted output $\hat{y} \in \mathbb{R}^{B \times H W S}$ is computed as:
\begin{align}
\hat{y} = \mathcal{M}(x, g, c),\, 
\label{network_output}
\end{align}
\noindent where $x$ is the input CT scan, $s$ is the input clinical data, and $c$ is  the center flag.  

\begin{figure*}[!bt]
\centering
\includegraphics[width=1\linewidth]{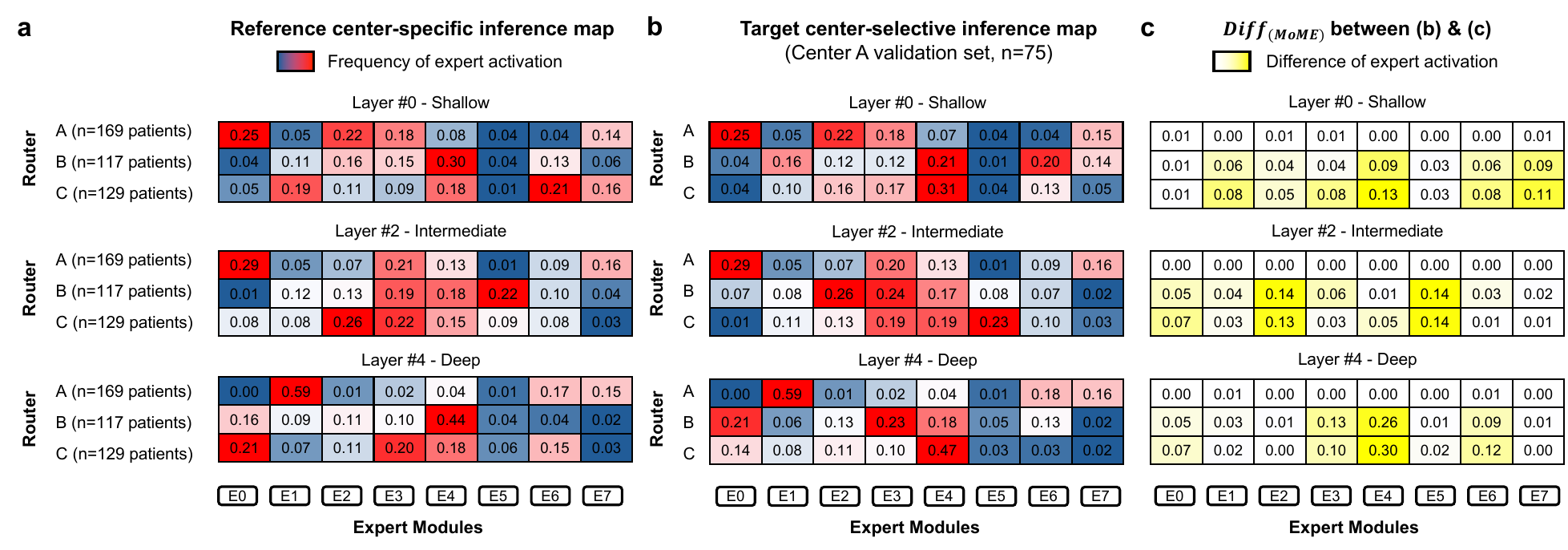}%
\caption{{Visualization of statistical analysis process for measuring center similarity. (a) Visualization of average frequency of activated top-$k$ experts for each reference dataset with corresponding center-specific router. (b) Visualization of frequency of activated top-$k$ experts for an independent Center A validation dataset. (c) Absolute difference of inference map between reference datasets in (a) and target dataset in (b), showing statistical center similarity $Diff_{(MoME)}$. }}
\label{fig_diff}
\end{figure*}

\vspace{10pt}
\subsubsection{Center-specific MoME inference} 
To perform inference using the trained MoME network on data from existing centers, we conduct center-specific inference based on the center flag $c\in \{A, B, C\}$. \add{For the closed center inference, we propose a statistical center similarity measure for selecting the optimal center flag $\hat{c}$ among previously involved training centers. During inference, the center-specific router automatically assigns the most appropriate expert layers for each input within the MoME module. To statistically measure the model's adaptation to each center, we count the number of times each expert was selected for every patch within the input and normalized these counts by the total number of patches, which are visualized in Fig.~\ref{fig_diff}(a). Next, {we provide a basis example for explaining the activation-frequency computation using an independent validation dataset from Center A, which was not included in training nor inference map preparation}. We examine the activation frequency of the top-$k$ expert modules. As shown in Fig.~\ref{fig_diff}(b), the activation patterns of the router A exhibited frequency trends similar to those in Fig.~\ref{fig_diff}(a). However, when activating the router B or C for this dataset, the patterns became distinct from those in Fig.~\ref{fig_diff}(a). This difference is visualized in Fig.~\ref{fig_diff}(c), which shows the absolute difference between the reference map in Fig.~\ref{fig_diff}(a) and the target map in Fig.~\ref{fig_diff}(b). Smaller differences observed for router A indicate that the MoME module effectively captures the underlying data distribution as trained, demonstrating its ability to adapt activations to the characteristics of each center dataset. In in-depth analysis of router A in Fig.~\ref{fig_diff}(c), sum of absolute differences for each layer $l$ are $0.022 \pm 0.003$, $0.013\pm 0.001$, $0.007\pm 0.001$, $0.023\pm 0.003$, and $0.030\pm 0.02$ as $l$ increased, indicating that expert activations in the intermediate layer are most similar to the reference inference map. This observation motivates the formulation of the statistical center similarity measure $Diff_{\text{MoME}}$ and the selection of the pseudo center $\hat{c}$ for the closed center inference.}

\add{Following the statistical analysis process illustrated in Fig.~\ref{fig_diff}, we compute the sum of absolute differences in expert activation frequency between the target closed centers $c' \in {D, E}$ and the reference centers $c \in {A, B, C}$, and define $Diff_{\text{MoME}}$ as: }
\add{
\begin{align}
Diff_{(MoME)} = \sum_{l=1}^{L} \alpha^l \sum_{e=1}^{n} \big| Freq_{c,e}^l - Freq_{c',e}^l \big|,
\label{diff}
\end{align}
}\add{where $L$ is the number of layers, $n$ is the number of expert modules, and $Freq_{c,e}^l$ is the normalized activation frequency of expert $e$ in layer $l$ for center $c$. To reflect the observation that the differences are minimal in the intermediate layers but more pronounced in the shallow and deep layers as shown in Fig.~\ref{fig_diff}(c), we assign layer-wise weights $\alpha^l = \{0.01,\,0.1,\,1.0,\,0.1,\,0.01\}$. For calculating $Freq_{c}$ for reference map, we use entire testset from each reference center, and for $Freq_{c'}$, {we use 20 fine-tuning samples from each closed center. To justify that the chosen sample size does not materially affect routing decisions, we performed a sensitivity test by varying the number of test cases  in Supplementary Table III.} Then we choose the pseudo center $\hat{c}\in \{A, B, C\}$ that minimizes $Diff_{(MoME)}$ for each closed center:}
\add{
\begin{align}
\hat{c} = \arg\min_{c} Diff_{(MoME)}.
\label{argmin}
\end{align}}
\vspace{-5pt}
\subsubsection{Closed center MoME fine-tuning} 
{To further fine-tune the MoME network in the closed center setting, we train the model based on the selected optimal pre-trained model weights based on the calculated pseudo center $\hat{c}\in \{A, B, C\}$ as a starting point. For fine-tuning the model, we basically follow the center-specific MoME training approach by utilizing few-shot dataset from each closed center. For efficient transfer of the pre-trained knowledge, the image encoder and the text prompt parameters are kept frozen.}

\subsection{Implementation details}
\label{met_implement}
Full implementation details, including data preprocessing, network configuration, training hyperparameters, and statistical analysis for confidence intervals (CIs), are provided in Supplementary Section II.

\subsection{Evaluation} To quantitatively assess PTV delineation performance, we calculate the Dice coefficient (Dice) and Intersection over Union (IoU) for each patient's PTV delineation result. \add{To evaluate equity-scaled (ES) performance across centers, we adopt the ES-Dice metric following \cite{tianfairseg}:}
\add{\begin{equation}\label{eq_essp}
\text{ES-Dice} = \frac{Dice(\hat{y}, y)}{1 + \Delta}, \
\end{equation}
\begin{equation}
\Delta = \sum_{c \in \{A,B,C\}} \left| Dice(\hat{y}, y) - Dice(\hat{y}, c, y) \right|,
\end{equation}}where $Dice(\hat{y}, y)$ denotes the Dice score computed across all centers jointly, and $Dice(\hat{y}, c, y)$ denotes the Dice score evaluated for each individual center $c \in \{A, B, C\}$. 
We further calculate the 95th percentile of the Hausdorff Distance (HD-95) \cite{crum2006generalized} to evaluate spatial discrepancies between the ground-truth and predicted contours. For reporting HD-95, all measured distances in pixel units are adjusted according to the original pixel resolution and reported in centimeters (cm). To further verify that the proposed MoME accurately reflects institutional characteristics, we evaluate inter-institutional PTV delineation patterns by using the Sacrum-to-PTV Ratio (SPR), defined as the ratio of total sacrum volume to total PTV volume. T
Sacrum labels for each CT scan were generated with the publicly available TotalSegmentator \cite{doi:10.1148/ryai.230024}. To ensure a consistent comparison across institutions, we exclusively analyze N0 patients, who have been pathologically diagnosed with no lymph node metastasis, and exclude N1 patients, whose treatment planning strategies may overlap across centers.

\begin{figure*}[!t]
\centering
\includegraphics[width=0.9\linewidth]{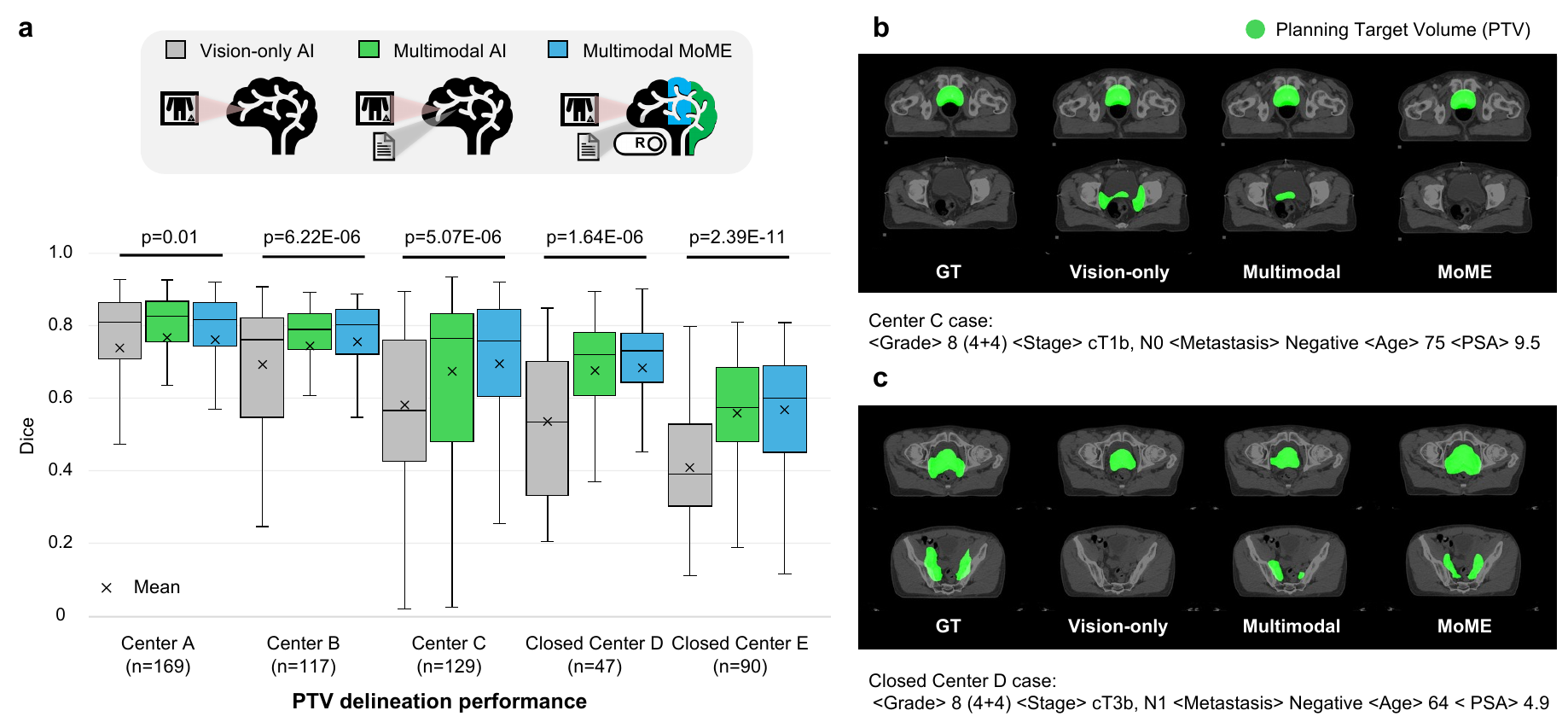}%
\caption{{Multicenter AI training comparison. (a) The multimodal MoME consistently achieves superior performance over vision-only and multimodal AI approaches. (b) In an intermediate-risk N0 patient case, the institution typically does not perform prophylactic nodal irradiation, yet both baselines erroneously included nodes in the delineation. In contrast, the MoME correctly focuses on the prostate, excluding the nodes. (c) In another intermediate-risk N1 patient case, the MoME delineates PTV with a larger margin, consistent with institutional practice, while both vision-only and multimodal AI approaches apply smaller margins. 
}}
\label{fig_internal}
\end{figure*}

\section{Experimental Results}



\subsection{Analysis of multicenter AI training performance}

We began by training baseline models under a traditional single-center AI training paradigm in Table~\ref{tab_main}(a). The vision-only model trained exclusively on data from Center A demonstrated overfitting to the training distribution, resulting in suboptimal performance on datasets from Centers B and C, which were not included in the training data. This yielded Dice scores of 0.681 and 0.559 for Centers B and C, respectively. Next, the multimodal AI approach incorporating both imaging and textual data, showed improved performance compared with the vision-only AI, with Dice scores of 0.739 and 0.633 for Centers B and C, respectively.

Next, we conducted experiments using the newly proposed multicenter AI training paradigm in Table~\ref{tab_main}(b), which included a few-shot datasets from Centers B and C alongside training data from Center A. All reported metrics represent results from the 1-shot setting. The vision-only AI exhibited comparable performance across multiple centers relative to single-center training. {Furthermore, federated learning approaches (FedAVG \cite{mcmahan2023fedavg}, FedProx \cite{li2020fedprox}) that aggregate model updates from decentralized multicenter data did not yield meaningful improvements over the LLMSeg baseline in the multicenter setting. Similarly, a contemporary multi-distribution fairness-oriented technique (FEBS \cite{tianfairseg}) and a multi-expert strategy for addressing inter-environment variability (MEDRO \cite{jeong2026multiexpert}) also failed to consistently surpass LLMSeg across the three centers. We further adapted the core concept of a recent group distributionally robust optimization method, PromptGDRO \cite{panaganti2026groupd}, to our center-specific distribution setting, but the results likewise remained inferior. In contrast, integrating our proposed MoME modules into the multimodal AI framework further improved PTV delineation performance, achieving Dice scores of 0.756, 0.752, and 0.692 for Center~A, B, and C, respectively.}

\add{However, the performance gain for Center A was not significant and was in some cases reduced when using the MoME module, as the previously overfitted model predictions were redistributed across strategies to better accommodate other centers. Nevertheless, the best ES-Dice score of 0.680 demonstrated equitable, debiased performance across centers when using the MoME module.} The performance gap and statistical significance among the vision-only AI, the multimodal AI (LLMSeg), and our proposed multimodal MoME for each center are further illustrated in the bar graph in Fig.~\ref{fig_internal}(a). We also performed qualitative comparisons of different approaches in the multicenter AI training setting to assess their clinical performance in Fig.~\ref{fig_internal}(b) and (c).

 
%
%
%

\begin{figure*}[!t]
\centering
\includegraphics[width=1\linewidth]{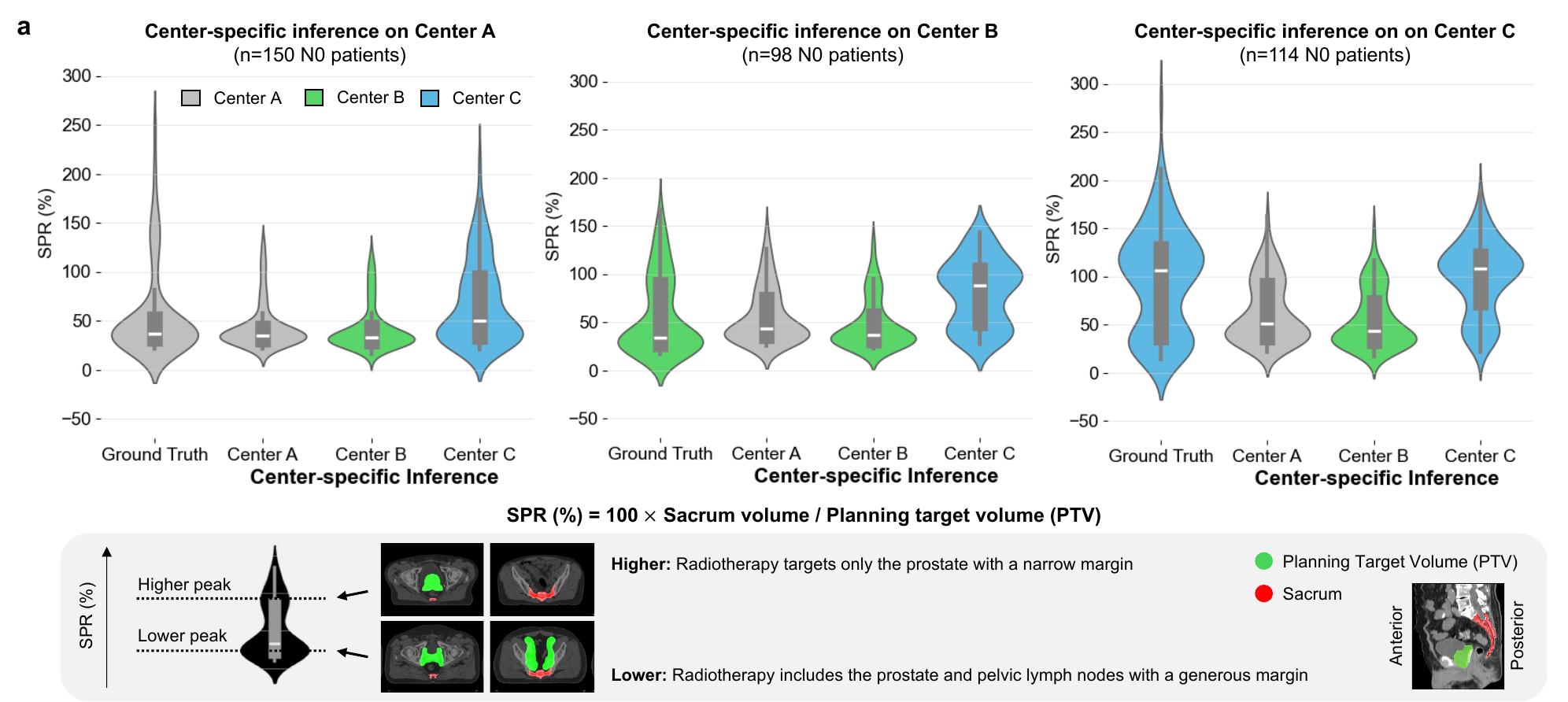}%
\caption{{Center-specific inference with the proposed MoME model demonstrates that radiotherapy planning strategies vary across institutions, as reflected in the Sacrum-to-PTV Ratio (SPR) distributions. Notably, the inferred distribution that best matches the ground truth for a given center (indicated by the same color) largely reflects its characteristic PTV distribution.}}
\label{fig_inference}
\end{figure*}

\begin{table*}[!t]
\centering
\caption{{PTV delineation performance for prostate cancer patients.
}}
\resizebox{1\linewidth}{!}{
\begin{tabular}{llcccccccccc}
\toprule

\multirow{3}{*}{\bf{Dataset}}
& \multirow{3}{*}{\bf{Metric}}
& \multicolumn{2}{c}{\bf{(a) Single-center AI Training}} & \multicolumn{8}{c}{\bf{(b) Multicenter AI Training}} \\

\cmidrule(l){3-4} \cmidrule(l){5-12}

& & \multirow{2}{*}{{Vision-only AI\cite{cciccek20163d}}} & \multicolumn{1}{c}{{Multimodal AI}}
& \multirow{2}{*}{{Vision-only AI\cite{cciccek20163d}}}
& \multicolumn{7}{c}{{Multimodal AI}}
\\
\cmidrule(l){4-4} \cmidrule(l){6-12}
&
&
& \multicolumn{1}{c}{{LLMSeg\cite{oh2024llm}}}
&
& \multicolumn{1}{c}{{LLMSeg\cite{oh2024llm}}}
& \multicolumn{1}{c}{{+FedAVG\cite{mcmahan2023fedavg}}}
& \multicolumn{1}{c}{{+FedProx\cite{li2020fedprox}}}
& \multicolumn{1}{c}{{+FEBS \cite{tianfairseg}}}
& \multicolumn{1}{c}{{+MEDRO \cite{jeong2026multiexpert}}}
& \multicolumn{1}{c}{{+PromptGDRO \cite{panaganti2026groupd}}}
& \multicolumn{1}{c}{{MoME (Ours)}}
\\

\cmidrule(l){1-1} \cmidrule(l){2-2} \cmidrule(l){3-3} \cmidrule(l){4-5}  \cmidrule(l){6-12}
\multirow{4}{*}{\shortstack[l]{\bf{Center A}\\\bf{(n=169)}}}

        & \shortstack[c]{{Dice $\uparrow$}\\{ }}   & \shortstack[c]{{0.725}\\{(0.696-0.751)}} & \shortstack[c]{{0.756}\\{(0.731-0.780)}}  & \shortstack[c]{{0.738}\\{(0.711-0.764)}} & \shortstack[c]{\bf{0.763}\\{(0.738-0.785)}} & \shortstack[c]{{0.757}\\{(0.731-0.782)}} & \shortstack[c]{{0.744}\\{(0.717-0.768)}} & \shortstack[c]{{0.757}\\{(0.733-0.780)}} & \shortstack[c]{{0.761}\\{(0.736-0.784)}} & \shortstack[c]{{0.757}\\{(0.732-0.780)}} & \shortstack[c]{{0.756}\\{(0.731-0.778)}}  \\
         & \shortstack[c]{{IoU $\uparrow$}\\{ }}   & \shortstack[c]{{0.598}\\{(0.568-0.626)}} & \shortstack[c]{{0.633}\\{(0.605-0.658)}}  & \shortstack[c]{{0.612}\\{(0.583-0.641)}} & \shortstack[c]{\bf{0.640}\\{(0.613-0.664)}} & \shortstack[c]{{0.634}\\{(0.606-0.661)}} & \shortstack[c]{{0.618}\\{(0.589-0.645)}} & \shortstack[c]{{0.633}\\{(0.606-0.658)}} & \shortstack[c]{{0.639}\\{(0.612-0.663)}} & \shortstack[c]{{0.633}\\{(0.606-0.658)}} & \shortstack[c]{{0.631}\\{(0.604-0.654)}}  \\
         & \shortstack[c]{{HD-95 $\downarrow$}\\{ }}  & \shortstack[c]{{1.630}\\{(1.333-1.945)}} & \shortstack[c]{{1.358}\\{(1.074-1.653)}}  & \shortstack[c]{{1.382}\\{(1.111-1.650)}} & \shortstack[c]{\bf{1.201}\\{(1.002-1.417)}} & \shortstack[c]{{3.640}\\{(3.060-4.274)}} & \shortstack[c]{{4.036}\\{(3.404-4.672)}} & \shortstack[c]{{1.165}\\{(0.956-1.404)}} & \shortstack[c]{{3.339}\\{(2.731-3.989)}} & \shortstack[c]{{3.453}\\{(2.909-4.008)}} & \shortstack[c]{{1.421}\\{(1.148-1.722)}}  \\

\cmidrule(l){1-1} \cmidrule(l){2-2} \cmidrule(l){3-3} \cmidrule(l){4-5}  \cmidrule(l){6-12}
\multirow{4}{*}{\shortstack[l]{\bf{Center B}\\\bf{(n=117)}}}

         & \shortstack[c]{{Dice $\uparrow$}\\{ }}   & \shortstack[c]{{0.681}\\{(0.651-0.709)}} & \shortstack[c]{{0.739}\\{(0.718-0.759)}}  & \shortstack[c]{{0.675}\\{(0.646-0.704)}} & \shortstack[c]{{0.741}\\{(0.715-0.766)}} & \shortstack[c]{{0.735}\\{(0.709-0.759)}} & \shortstack[c]{{0.724}\\{(0.703-0.745)}} & \shortstack[c]{{0.722}\\{(0.694-0.749)}} & \shortstack[c]{{0.748}\\{(0.725-0.771)}} & \shortstack[c]{{0.727}\\{(0.700-0.752)}} & \shortstack[c]{\bf{0.752}\\{(0.729-0.773)}}  \\
         & \shortstack[c]{{IoU $\uparrow$}\\{ }}   & \shortstack[c]{{0.535}\\{(0.503-0.565)}} & \shortstack[c]{{0.598}\\{(0.575-0.621)}}  & \shortstack[c]{{0.527}\\{(0.496-0.559)}} & \shortstack[c]{{0.605}\\{(0.577-0.632)}} & \shortstack[c]{{0.596}\\{(0.569-0.624)}} & \shortstack[c]{{0.580}\\{(0.557-0.603)}} & \shortstack[c]{{0.583}\\{(0.553-0.612)}} & \shortstack[c]{{0.612}\\{(0.586-0.637)}} & \shortstack[c]{{0.587}\\{(0.558-0.615)}} & \shortstack[c]{\bf{0.616}\\{(0.590-0.640)}}  \\
         & \shortstack[c]{{HD-95 $\downarrow$}\\{ }}  & \shortstack[c]{{1.741}\\{(1.541-1.950)}} & \shortstack[c]{{1.384}\\{(1.219-1.558)}}  & \shortstack[c]{{1.723}\\{(1.497-1.943)}} & \shortstack[c]{\bf{1.247}\\{(1.073-1.429)}} & \shortstack[c]{{3.122}\\{(2.821-3.438)}} & \shortstack[c]{{3.157}\\{(2.925-3.386)}} & \shortstack[c]{{1.444}\\{(1.229-1.679)}} & \shortstack[c]{{2.949}\\{(2.687-3.207)}} & \shortstack[c]{{3.247}\\{(2.915-3.607)}} & \shortstack[c]{{1.331}\\{(1.141-1.511)}}  \\

\cmidrule(l){1-1} \cmidrule(l){2-2} \cmidrule(l){3-3} \cmidrule(l){4-5}  \cmidrule(l){6-12}
\multirow{4}{*}{\shortstack[l]{\bf{Center C}\\\bf{(n={129})}}}

        & \shortstack[c]{{Dice $\uparrow$}\\{ }}   & \shortstack[c]{{0.559}\\{(0.526-0.595)}} & \shortstack[c]{{0.633}\\{(0.597-0.670)}}  & \shortstack[c]{{0.563}\\{(0.529-0.600)}} & \shortstack[c]{{0.671}\\{(0.635-0.708)}} & \shortstack[c]{{0.606}\\{(0.567-0.642)}} & \shortstack[c]{{0.609}\\{(0.577-0.641)}} & \shortstack[c]{{0.675}\\{(0.640-0.711)}} & \shortstack[c]{{0.628}\\{(0.591-0.663)}} & \shortstack[c]{{0.646}\\{(0.609-0.679)}} & \shortstack[c]{\bf{0.692}\\{(0.661-0.725)}}  \\
         & \shortstack[c]{{IoU $\uparrow$}\\{ }}   & \shortstack[c]{{0.412}\\{(0.380-0.447)}} & \shortstack[c]{{0.494}\\{(0.457-0.533)}}  & \shortstack[c]{{0.418}\\{(0.383-0.455)}} & \shortstack[c]{{0.536}\\{(0.497-0.574)}} & \shortstack[c]{{0.466}\\{(0.428-0.504)}} & \shortstack[c]{{0.463}\\{(0.431-0.497)}} & \shortstack[c]{{0.540}\\{(0.501-0.578)}} & \shortstack[c]{{0.489}\\{(0.451-0.525)}} & \shortstack[c]{{0.507}\\{(0.470-0.543)}} & \shortstack[c]{\bf{0.557}\\{(0.525-0.591)}}  \\
         & \shortstack[c]{{HD-95 $\downarrow$}\\{ }}  & \shortstack[c]{{2.756}\\{(2.463-3.041)}} & \shortstack[c]{{2.473}\\{(2.071-2.967)}} & \shortstack[c]{{2.629}\\{(2.364-2.893)}} & \shortstack[c]{{1.949}\\{(1.654-2.298)}} & \shortstack[c]{{6.964}\\{(5.810-8.207)}} & \shortstack[c]{{6.542}\\{(5.471-7.619)}} & \shortstack[c]{\bf{1.926}\\{(1.589-2.307)}} & \shortstack[c]{{5.850}\\{(4.796-7.104)}} & \shortstack[c]{{5.510}\\{(4.640-6.398)}} & \shortstack[c]{{2.395}\\{(1.980-2.897)}} \\

\cmidrule(l){1-1} \cmidrule(l){2-2} \cmidrule(l){3-3} \cmidrule(l){4-5}  \cmidrule(l){6-12}
\shortstack[l]{\bf{All (n={415})}\\{}}  & \shortstack[c]{{ES-Dice $\uparrow$}\\{}} & \shortstack[c]{{0.556}\\{(0.536-0.578)}} & \shortstack[c]{{0.621}\\{(0.596-0.647)}} & \shortstack[c]{{0.560}\\{(0.540-0.581)}} & \shortstack[c]{{0.658}\\{(0.634-0.685)}} & \shortstack[c]{{0.595}\\{(0.570-0.619)}} & \shortstack[c]{{0.598}\\{(0.575-0.620)}} & \shortstack[c]{{0.660}\\{(0.633-0.687)}} & \shortstack[c]{{0.614}\\{(0.589-0.639)}} & \shortstack[c]{{0.633}\\{(0.609-0.657)}} & \shortstack[c]{\bf{0.680}\\{(0.652-0.707)}} \\

\bottomrule
\multicolumn{12}{l}{{{Note.} {\textbf{Bold} metric indicates best performance. All reported metrics in (b) are obtained from the 1-shot setting. ES-Dice evaluates debiased performance across three centers.}}}

\end{tabular}
}
\label{tab_main}
\end{table*}

\begin{table*}[!h]
\caption{{PTV delineation performance on closed center dataset with different size of few-shot fine-tuning dataset.}}
\centering
\resizebox{1\linewidth}{!}{
\begin{tabular}{llcccccccccccc}

\toprule

\multirow{2}{*}{\bf{Method}}
& \multirow{2}{*}{\bf{Metric}}
& \multicolumn{6}{c}{\bf{Closed Center D (n=47)}} & \multicolumn{6}{c}{\bf{Closed Center E (n=90)}} \\
\cmidrule(l){3-8}  \cmidrule(l){9-14}

& & \multicolumn{3}{c}{{Inference without fine-tuning (0-shot)}}
& \multicolumn{1}{c}{{1-shot}}
& \multicolumn{1}{c}{{2-shots}}
& \multicolumn{1}{c}{{3-shots}}
& \multicolumn{3}{c}{{Inference without fine-tuning (0-shot)}}
& \multicolumn{1}{c}{{1-shot}}
& \multicolumn{1}{c}{{2-shots}}
& \multicolumn{1}{c}{{3-shots}}
\\

\cmidrule(l){1-1} \cmidrule(l){2-2} \cmidrule(l){3-8}   \cmidrule(l){9-14} 

    \multirow{4}{*}{\shortstack[l]{\bf{Vision-only} \\ \bf{AI}}} 
    & \shortstack[c]{{Dice $\uparrow$}\\{ }}   
    & \multicolumn{3}{c}{\shortstack[c]{{0.384}\\{(0.320-0.448)}}}  & \shortstack[c]{{0.401}\\{(0.334-0.468)}}  & \shortstack[c]{{0.420}\\{(0.360-0.476)}} & \shortstack[c]{{0.473}\\{(0.433-0.513)}}  & \multicolumn{3}{c}{\shortstack[c]{{0.347}\\{(0.314-0.380)}}}  & \shortstack[c]{{0.413}\\{(0.378-0.447)}} &  \shortstack[c]{{0.471}\\{(0.436-0.505)}}  & \shortstack[c]{{0.463}\\{(0.428-0.495)}}  \\ 

    & \shortstack[c]{{IoU $\uparrow$}\\{ }}   & \multicolumn{3}{c}{\shortstack[c]{{0.263}\\{(0.211-0.316)}}}  & \shortstack[c]{{0.279}\\{(0.223-0.334)}}  &  \shortstack[c]{{0.288}\\{(0.238-0.336)}}  & \shortstack[c]{{0.321}\\{(0.287-0.355)}}  
     & \multicolumn{3}{c}{\shortstack[c]{{0.223}\\{(0.196-0.249)}}} &  \shortstack[c]{{0.273}\\{(0.245-0.301)}} & \shortstack[c]{{0.323}\\{(0.292-0.353)}} & \shortstack[c]{{0.315}\\{(0.286-0.344)}}  \\ 
 
    & \shortstack[c]{{HD-95 $\downarrow$}\\{ }}  & \multicolumn{3}{c}{\shortstack[c]{{6.507}\\{(5.577-7.511)}}}  & \shortstack[c]{{6.651}\\{(5.712-7.646)}}  & \shortstack[c]{{6.706}\\{(5.760-7.762)}}  & \shortstack[c]{{6.119}\\{(5.197-7.170)}}   & \multicolumn{3}{c}{\shortstack[c]{{4.050}\\{(3.590-4.615)}}}  &  \shortstack[c]{{3.955}\\{(3.390-4.563)}} &  \shortstack[c]{{2.810}\\{(2.469-3.232)}} & \shortstack[c]{{4.188}\\{(3.766-4.701)}} \\ 
    
\cmidrule(l){1-1} \cmidrule(l){2-2} \cmidrule(l){3-5} \cmidrule(l){6-6} \cmidrule(l){7-7} \cmidrule(l){8-8}  \cmidrule(l){9-11} \cmidrule(l){12-12} \cmidrule(l){13-13} \cmidrule(l){14-14}  

\multirow{4}{*}{\shortstack[l]{\bf{Multimodal} \\ \bf{AI}}} & 
    \shortstack[c]{{Dice $\uparrow$}\\{ }}  & \multicolumn{3}{c}{\shortstack[c]{{0.568}\\{(0.521-0.613)}}} & \shortstack[c]{{0.610}\\{(0.552-0.662)}} & \shortstack[c]{{0.656}\\{(0.606-0.702)}} & \shortstack[c]{{0.673}\\{(0.627-0.717)}}
    & \multicolumn{3}{c}{\shortstack[c]{{0.411}\\{(0.377-0.446)}}}  & \shortstack[c]{{0.559}\\{(0.526-0.591)}}  & \shortstack[c]{\bf{0.604}\\{(0.568-0.641)}} & \shortstack[c]{{0.610}\\{(0.571-0.643)}}   \\ 

    & \shortstack[c]{{IoU $\uparrow$}\\{ }}   & \multicolumn{3}{c}{ \shortstack[c]{{0.412}\\{(0.370-0.451)}}} & \shortstack[c]{{0.462}\\{(0.410-0.509)}} & \shortstack[c]{{0.507}\\{(0.460-0.552)}} & \shortstack[c]{{0.524}\\{(0.478-0.569)}}  & 
    \multicolumn{3}{c}{\shortstack[c]{{0.274}\\{(0.245-0.304)}}}  & \shortstack[c]{{0.404}\\{(0.373-0.434)}} &  \shortstack[c]{\bf{0.449}\\{(0.414-0.485)}}  & \shortstack[c]{{0.458}\\{(0.423-0.489)}}  \\

    & \shortstack[c]{{HD-95 $\downarrow$}\\{ }}  & \multicolumn{3}{c}{\shortstack[c]{{6.672}\\{(5.347-8.142)}}} & \shortstack[c]{{8.173}\\{(6.283-10.041)}} & \shortstack[c]{{5.567}\\{(4.146-7.152)}} & \shortstack[c]{{4.595}\\{(3.360-5.911)}}  & 
    \multicolumn{3}{c}{\shortstack[c]{{3.419}\\{(3.024-3.832)}}}  &  \shortstack[c]{\bf{1.890}\\{(1.529-2.308)}} & \shortstack[c]{{4.299}\\{(3.367-5.405)}} &  \shortstack[c]{\underline{\bf{1.737}}\\{(1.366-2.181)}}  \\ 

\cmidrule(l){1-1} \cmidrule(l){2-2} \cmidrule(l){3-5} \cmidrule(l){6-6} \cmidrule(l){7-7} \cmidrule(l){8-8}  \cmidrule(l){9-11} \cmidrule(l){12-12} \cmidrule(l){13-13} \cmidrule(l){14-14}  

\multirow{11}{*}{\shortstack[l]{\bf{MoME} \\ \bf{(Ours)}} } &  & \multicolumn{3}{c}{Center-specific Inference}  & & & & \multicolumn{3}{c}{Center-specific Inference} & & & \\
&  &  {Center A} & {Center B} & {Center C}  &  & & &  {Center A} & {Center B} & {Center C}  & & & \\
\cmidrule(l){2-2} \cmidrule(l){3-5} \cmidrule(l){6-6} \cmidrule(l){7-7} \cmidrule(l){8-8}  \cmidrule(l){9-11} \cmidrule(l){12-12} \cmidrule(l){13-13} \cmidrule(l){14-14}  

    & \shortstack[c]{{Dice $\uparrow$}\\{ }}   & \shortstack[c]{{0.585}\\{(0.550-0.618)}} & \shortstack[c]{{0.548}\\{(0.506-0.594)}} & \shortstack[c]{\bf{0.605}\\{(0.577-0.629)}} & \shortstack[c]{\bf{0.628}\\{(0.581-0.673)}} & \shortstack[c]{\underline{\bf{0.682}}\\{(0.642-0.722)}} & \shortstack[c]{\bf{0.677}\\{(0.637-0.716)}}  & 
    
    \shortstack[c]{{0.393}\\{(0.358-0.425)}} &  \shortstack[c]{{0.406}\\{(0.371-0.438)}} &  \shortstack[c]{\bf{0.490}\\{(0.460-0.522)}}  & \shortstack[c]{\bf{0.568}\\{(0.534-0.600)}}  & \shortstack[c]{{0.596}\\{(0.563-0.626)}}  & \shortstack[c]{\underline{\bf{0.612}}\\{(0.573-0.646)}}  \\ 

    & \shortstack[c]{{IoU $\uparrow$}\\{ }}   & \shortstack[c]{{0.422}\\{(0.390-0.454)}} & \shortstack[c]{{0.393}\\{(0.353-0.434)}} & \shortstack[c]{\bf{0.439}\\{(0.412-0.464)}} & \shortstack[c]{\bf{0.476}\\{(0.430-0.520)}} & \shortstack[c]{\underline{\bf{0.533}}\\{(0.491-0.577)}} & \shortstack[c]{\bf{0.526}\\{(0.485-0.568)}} &
    
    \shortstack[c]{{0.260}\\{(0.230-0.287)}}  &  \shortstack[c]{{0.270}\\{(0.240-0.298)}}  & \shortstack[c]{\bf{0.338}\\{(0.312-0.365)}} & \shortstack[c]{\bf{0.413}\\{(0.382-0.444)}}  & \shortstack[c]{{0.441}\\{(0.409-0.469)}} &  \shortstack[c]{\underline{\bf{0.461}}\\{(0.425-0.494)}} \\ 

    & \shortstack[c]{{HD-95 $\downarrow$}\\{ }}  & \shortstack[c]{\bf{5.767}\\{(4.596-7.121)}} & \shortstack[c]{{6.010}\\{(5.167-7.012)}} & \shortstack[c]{{6.369}\\{(5.002-7.792)}} & \shortstack[c]{\bf{6.606}\\{(5.285-7.995)}} & \shortstack[c]{\bf{5.161}\\{(3.782-6.641)}} & \shortstack[c]{\underline{\bf{4.283}}\\{(3.413-5.163)}}  &  
    
    \shortstack[c]{{3.365}\\{(2.980-3.807)}} & \shortstack[c]{{3.230}\\{(2.721-3.787)}} & \shortstack[c]{\bf{2.810}\\{(2.267-3.417)}} & \shortstack[c]{{2.049}\\{(1.635-2.550)}} &  \shortstack[c]{\bf{1.766}\\{(1.396-2.248)}} &  \shortstack[c]{{1.788}\\{(1.400-2.299)}} 
\\ 

 \cmidrule(l){2-2} \cmidrule(l){3-5} \cmidrule(l){6-6} \cmidrule(l){7-7} \cmidrule(l){8-8}  \cmidrule(l){9-11} \cmidrule(l){12-12} \cmidrule(l){13-13} \cmidrule(l){14-14}  

& $\bf{Diff_{(MoME)}} \downarrow$ & 0.85 & 0.53 & \bf{0.44} & & & & 0.77 & 0.56 & \bf{0.52} & & & \\

\bottomrule

\multicolumn{14}{l}{{{Note.} {\textbf{Bold} metric indicates best performance among different few-shot dataset settings, whereas,  \underline{underline} for among entire trials, for each center.}}}
     
\end{tabular}
}
\label{tab_finetune}
\end{table*}

\subsection{Center-specific inference reflects institutional strategy}

\add{During inference, a key advantage of our MoME module is its ability to select a center-specific router tailored to each center's dataset characteristics. This capability allows us to analyze model predictions by choosing a corresponding or different router path. To evaluate the overall tendencies of each center-specific router, we first tested the entire test dataset from each center as input, activating the corresponding center-specific routers. Specifically, we visualized the overall trends in how each router captures the center’s unique target delineation strategy using violin plots of SPR values, enabling us to assess how target distribution shifts with the application of these center-specific routers. As shown in Fig.~\ref{fig_inference}(a), the SPR distribution closely aligned with each center’s clinical practices when using the corresponding expert router, with Centers A and B exhibiting similar patterns characterized by frequent PNI and broader margins, while Center C showed distinct trends with higher SPR values due to less frequent PNI and tighter PTV margins. Risk group analysis further confirmed that Center-specific experts produced SPR distributions consistent with their respective institutional practices, with Centers A and B showing greater similarity compared to Center C, as detailed in Supplementary Fig. 1(a)-(f).}

\begin{table*}[!h]
    \caption{{Ablation studies on the network training strategy and architectural components.}}
    \centering
    \resizebox{1\linewidth}{!}{
    \begin{tabular}{llcccccccccccclccc}
    \toprule
         \multirow{2}{*}{\bf{Dataset}} & \multirow{2}{*}{\bf{Metric}} & \multirow{2}{*}{\shortstack[l]{{}\\\bf{MoME}\\\bf{(Ours)}}} & \multicolumn{2}{c}{\bf{(a) Multicenter Training Method}}  & \multicolumn{2}{c}{\bf{(b) Top-$k$ for MoME}} & \multicolumn{2}{c}{\bf{(c) Total Number (n) of Experts}} & \multicolumn{2}{c}{{\bf{(d) Expert Module Design}}} & \multicolumn{2}{c}{{\bf{(e) Choice of LLM}}} &  \multirow{2}{*}{\bf{Dataset}} & \multicolumn{3}{c}{\bf{\add{(f) Number of Training Centers}}} \\
        \cmidrule(l){4-5} \cmidrule(l){6-7} \cmidrule(l){8-9} \cmidrule(l){10-11} \cmidrule(l){12-13} \cmidrule(l){15-17}
          & & & Text Prompt  & Vanilla MoE & Top-1 & Top-3 & 4 Experts & 16 Experts & {Linear} & {Modern (SiLU)} & {CLIP} & {Qwen3-8B} & & 1 (A) & 2 (A,B$^f$) & 3 (A,B$^f$,C$^f$)\\
        \toprule

        \multirow{4}{*}{\shortstack[l]{\bf{Center A}\\\bf{(n=169)}}}  & \shortstack[c]{{Dice $\uparrow$}\\{ }}   & \shortstack[c]{{0.756}\\{(0.731-0.778)}} & \shortstack[c]{{0.735}\\{(0.708-0.762)}} & \shortstack[c]{{0.757}\\{(0.730-0.782)}} & \shortstack[c]{{0.753}\\{(0.726-0.778)}} & \shortstack[c]{{0.754}\\{(0.730-0.778)}} & \shortstack[c]{{0.756}\\{(0.731-0.779)}} & \shortstack[c]{{0.756}\\{(0.731-0.779)}} & {\shortstack[c]{0.742\\(0.716-0.767)}} & {\shortstack[c]{0.756\\(0.729-0.780)}} & {\shortstack[c]{0.745\\(0.717-0.771)}} & {\shortstack[c]{0.745\\(0.717-0.770)}} \\
         & \shortstack[c]{{IoU $\uparrow$}\\{ }}   & \shortstack[c]{{0.631}\\{(0.604-0.654)}} & \shortstack[c]{{0.609}\\{(0.580-0.636)}} & \shortstack[c]{{0.635}\\{(0.605-0.659)}} & \shortstack[c]{{0.631}\\{(0.603-0.659)}} & \shortstack[c]{{0.631}\\{(0.604-0.656)}} & \shortstack[c]{{0.632}\\{(0.605-0.657)}} & \shortstack[c]{{0.632}\\{(0.605-0.657)}} & {\shortstack[c]{0.616\\(0.589-0.643)}} & {\shortstack[c]{\bf{0.633}\\(0.603-0.659)}} & {\shortstack[c]{0.622\\(0.593-0.651)}} & {\shortstack[c]{0.622\\(0.592-0.649)}} \\

         & \shortstack[c]{{HD-95 $\downarrow$}\\{ }} & \shortstack[c]{{1.421}\\{(1.148-1.722)}} & \shortstack[c]{{1.444}\\{(1.177-1.725)}} & \shortstack[c]{{1.390}\\{(1.079-1.727)}} & \shortstack[c]{{1.507}\\{(1.182-1.844)}} & \shortstack[c]{{1.290}\\{(1.062-1.535)}} & \shortstack[c]{{1.353}\\{(1.052-1.669)}} & \shortstack[c]{{1.353}\\{(1.052-1.669)}} & {\shortstack[c]{3.933\\(3.415-4.479)}} & {\shortstack[c]{3.174\\(2.734-3.625)}} & {\shortstack[c]{3.263\\(2.823-3.729)}} & {\shortstack[c]{3.246\\(2.785-3.721)}} \\

        \cmidrule(l){1-1} \cmidrule(l){2-2} \cmidrule(l){3-3} \cmidrule(l){4-5} \cmidrule(l){6-7} \cmidrule(l){8-9} \cmidrule(l){10-11} \cmidrule(l){12-13} \cmidrule(l){14-14} \cmidrule(l){15-17}

        \multirow{4}{*}{\shortstack[l]{\bf{Center B}\\\bf{(n=117)}}} & \shortstack[c]{{Dice $\uparrow$}\\{ }}   & \shortstack[c]{{0.752}\\{(0.729-0.773)}} & \shortstack[c]{{0.740}\\{(0.715-0.764)}} & \shortstack[c]{{0.748}\\{(0.724-0.770)}} & \shortstack[c]{{0.755}\\{(0.731-0.777)}} & \shortstack[c]{{0.739}\\{(0.710-0.765)}} & \shortstack[c]{{0.760}\\{(0.734-0.782)}} & \shortstack[c]{{0.740}\\{(0.712-0.766)}} & {\shortstack[c]{0.740\\(0.714-0.765)}} & {\shortstack[c]{{0.763}\\(0.737-0.787)}} & {\shortstack[c]{0.717\\(0.684-0.746)}} & {\shortstack[c]{{0.754}\\(0.726-0.780)}}  & \multirow{4}{*}{\shortstack[l]{\bf{Closed} \\ \bf{Center D}\\\bf{(n=47)}}} & \shortstack[c]{{0.577}\\{(0.535-0.618)}} & \shortstack[c]{{0.550}\\{(0.514-0.586)}} & \shortstack[c]{{0.605}\\{(0.577-0.629)}} \\

         & \shortstack[c]{{IoU $\uparrow$}\\{ }}   & \shortstack[c]{{0.616}\\{(0.590-0.640)}} & \shortstack[c]{{0.603}\\{(0.575-0.629)}} & \shortstack[c]{{0.612}\\{(0.585-0.637)}} & \shortstack[c]{{0.622}\\{(0.595-0.647)}} & \shortstack[c]{{0.607}\\{(0.575-0.636)}} & \shortstack[c]{{0.629}\\{(0.600-0.654)}} & \shortstack[c]{{0.607}\\{(0.577-0.637)}} & {\shortstack[c]{0.604\\(0.575-0.632)}} & {\shortstack[c]{{0.634}\\(0.606-0.661)}} & {\shortstack[c]{0.583\\(0.546-0.616)}} & {\shortstack[c]{{0.624}\\(0.594-0.653)}} & & \shortstack[c]{{0.419}\\{(0.380-0.457)}} & \shortstack[c]{{0.389}\\{(0.357-0.422)}} & \shortstack[c]{{0.439}\\{(0.412-0.464)}} \\

         & \shortstack[c]{{HD-95 $\downarrow$}\\{ }}  & \shortstack[c]{{1.331}\\{(1.141-1.511)}} & \shortstack[c]{{1.369}\\{(1.181-1.556)}} & \shortstack[c]{{1.273}\\{(1.107-1.452)}} & \shortstack[c]{{1.257}\\{(1.078-1.446)}} & \shortstack[c]{{1.310}\\{(1.110-1.531)}} & \shortstack[c]{{1.377}\\{(1.181-1.584)}} & \shortstack[c]{{1.306}\\{(1.108-1.528)}} & {\shortstack[c]{3.411\\(3.084-3.753)}} & {\shortstack[c]{2.674\\(2.396-2.963)}} & {\shortstack[c]{3.389\\(3.076-3.716)}} & {\shortstack[c]{2.885\\(2.598-3.209)}} & & \shortstack[c]{{5.808}\\{(4.398-7.294)}} & \shortstack[c]{{8.398}\\{(6.829-9.967)}}  &\shortstack[c]{{6.369}\\{(5.002-7.792)}}\\

        \cmidrule(l){1-1} \cmidrule(l){2-2} \cmidrule(l){3-3} \cmidrule(l){4-5} \cmidrule(l){6-7} \cmidrule(l){8-9} \cmidrule(l){10-11} \cmidrule(l){12-13} \cmidrule(l){14-14} \cmidrule(l){15-17}

        \multirow{4}{*}{\shortstack[l]{\bf{Center C}\\\bf{(n={129})}}} & \shortstack[c]{{Dice $\uparrow$}\\{ }}   & \shortstack[c]{{0.692}\\{(0.661-0.725)}} & \shortstack[c]{{0.627}\\{(0.590-0.664)}} & \shortstack[c]{{0.650}\\{(0.614-0.687)}} & \shortstack[c]{{0.679}\\{(0.648-0.710)}} & \shortstack[c]{{0.694}\\{(0.659-0.727)}} & \shortstack[c]{{0.685}\\{(0.652-0.718)}} & \shortstack[c]{{0.694}\\{(0.663-0.727)}} & {\shortstack[c]{0.663\\(0.632-0.695)}} & {\shortstack[c]{0.690\\(0.655-0.720)}} & {\shortstack[c]{0.634\\(0.600-0.667)}} & {\shortstack[c]{0.665\\(0.631-0.697)}}  & \multirow{4}{*}{\shortstack[l]{\bf{Closed} \\ \bf{Center E}\\\bf{(n=90)}}} & \shortstack[c]{{0.477}\\{(0.441-0.512)}} & \shortstack[c]{{0.444}\\{(0.411-0.479)}} & \shortstack[c]{{0.490}\\{(0.460-0.522)}}\\

         & \shortstack[c]{{IoU $\uparrow$}\\{ }}   & \shortstack[c]{{0.557}\\{(0.525-0.591)}} & \shortstack[c]{{0.487}\\{(0.449-0.525)}} & \shortstack[c]{{0.513}\\{(0.475-0.551)}} & \shortstack[c]{{0.542}\\{(0.509-0.575)}} & \shortstack[c]{{0.562}\\{(0.526-0.597)}} & \shortstack[c]{{0.549}\\{(0.514-0.584)}} & \shortstack[c]{{0.560}\\{(0.527-0.595)}} & {\shortstack[c]{0.522\\(0.489-0.555)}} & {\shortstack[c]{0.554\\(0.519-0.587)}} & {\shortstack[c]{0.495\\(0.459-0.530)}} & {\shortstack[c]{0.525\\(0.489-0.559)}} & & \shortstack[c]{{0.330}\\{(0.298-0.362)}} & \shortstack[c]{{0.300}\\{(0.271-0.329)}} & \shortstack[c]{{0.338}\\{(0.312-0.365)}} \\

         & \shortstack[c]{{HD-95 $\downarrow$}\\{ }}  & \shortstack[c]{{2.395}\\{(1.980-2.897)}} & \shortstack[c]{{2.346}\\{(1.979-2.707)}} & \shortstack[c]{{2.296}\\{(1.942-2.655)}} & \shortstack[c]{{2.601}\\{(2.119-3.067)}} & \shortstack[c]{{2.444}\\{(1.907-3.030)}} & \shortstack[c]{{2.005}\\{(1.591-2.433)}} & \shortstack[c]{{2.480}\\{(1.949-3.021)}} & {\shortstack[c]{9.225\\(8.127-10.373)}} & {\shortstack[c]{8.799\\(7.344-10.349)}} & {\shortstack[c]{7.292\\(6.123-8.511)}} & {\shortstack[c]{6.324\\(5.266-7.469)}} & & \shortstack[c]{{3.061}\\{(2.485-3.753)}} & \shortstack[c]{{3.232}\\{(2.800-3.671)}} & \shortstack[c]{{2.810}\\{(2.267-3.417)}}\\

         \bottomrule
        \multicolumn{17}{l}{{{Note.} {Default MoME uses MLP expert, $k$=2, $n$=8, LLaMA3-Instruct-8B as LLM, with 3 training centers involved, where $^f$ indicates few-shots. All reported metrics for (a-e) represent results from the 1-shot setting, while 0-shot inference results for (f).}}}
    \end{tabular}
    }
\label{tab_ablation}
\end{table*}

\begin{table}[!t]
\centering
\caption{{Computational cost comparison.}}
\resizebox{1\linewidth}{!}{
\begin{tabular}{lcccc}
\toprule

\multirow{2}{*}{\bf{Metric}} 
& \multicolumn{1}{c}{\bf{Vision-only AI}} & \multicolumn{3}{c}{\bf{Multimodal AI}} \\

\cmidrule(l){2-2} \cmidrule(l){3-5}  
& \multirow{1}{*}{{3D ResUNet\cite{cciccek20163d}}} &   
\multicolumn{1}{c}{{3D LLMSeg \cite{oh2024llm}}} & Vanilla MoE & \multicolumn{1}{c}{{MoME (Ours)}} \\

\midrule  
\shortstack[c]{Network parameters} & 13.28 M & 34.48 M & 34.54 M & 34.54 M  \\ 
\shortstack[c]{FLOPs}   & 1542.36 G & 2.44 T  & 2.50 T  & 2.50 T \\ 
\shortstack[c]{Inference latency (s)}  & 1.162 $\pm$ 0.158 & 0.958 $\pm$ 0.440 & 1.479 $\pm$ 0.672 & 1.458 $\pm$ 0.660 \\ 
 
\bottomrule
\end{tabular}
}
\label{tab_cost}
\end{table}

\subsection{Data efficient few-shot fine-tuning on closed center dataset}


For the closed center setting, we monitored the performance of each multicenter AI training method as the size of the few-shot fine-tuning dataset progressively increased. Table~\ref{tab_finetune} summarizes the performance across diverse closed center datasets. For Center D, in the 0-shot inference setting, both baseline models showed suboptimal performance. In contrast, the MoME approach, which leveraged Center C-specific inference based on the $Diff_{(MoME)}$ measure, achieved performance improvements of up to 22\% over the vision-only AI. For Center E, in the 0-shot inference setting, both baseline models exhibited limited effectiveness with Dice score of around 0.400. In contrast, the MoME approach, utilizing Center C-specific routers based on the minimal $Diff_{(MoME)}$ score, achieved performance improvements of up to 14\% over the vision-only AI. This improvement is notable because Center C shares the most similar data acquisition conditions with both Center D and E, as analyzed in Fig.~\ref{fig_intro}(a). 

During subsequent few-shot fine-tuning from the selected pre-trained checkpoint, our MoME consistently enhanced the performance of multimodal baselines across all fine-tuning settings for Center D and achieved the best performance for Center E, as illustrated in Supplementary Fig.~2(a) and (d), respectively. To capture the richer prediction distribution, we further analyzed the SPR distributions under the closed center setting, as shown in Supplementary Fig.~2(b)-(c) and (e)–(f), for Centers D and E, respectively. During zero-shot inference, the vision-only AI consistently skewed toward lower SPR values, failing to adequately capture the clinical context for both centers. The multimodal AI similarly produced SPR distributions that deviated substantially from the ground truth. While the MoME outperformed both vision-only and multimodal AIs, notable discrepancies remained relative to the ground truth SPR distributions. However, when fine-tuning was performed with limited few-shot sampled from closed center data, a clear trend of improvement emerged. As the number of fine-tuning samples increased, the SPR distributions progressively aligned more closely with the ground truth for MoME. This improvement was consistently observed further across all risk groups in Supplementary Figs.~3 and 4. 
{We further conducted in-depth analyses on the MoME fine-tuning, including comparison with a from-scratch baseline trained only on the closed-center few-shot data without external-center pretraining (Supplementary Table~IV) and validation Dice evolution across training epochs (Supplementary Fig.~5).}


\subsection{Ablation studies in MoME training strategy}

We conducted ablation studies to assess the contribution of MoME components in Table~\ref{tab_ablation}. For the multicenter training method (Table~\ref{tab_ablation}(a)), we compared MoME against (i) a Text Prompt baseline that appends the center title (e.g., {\em``Center C"}) to the clinical input within the multimodal AI framework, and (ii) a Vanilla MoE that uses a unified router across all centers without a center-specific router. MoME consistently surpasses the Text Prompt baseline, indicating that center-specific routing is more effective than textual center cues via a single path. Compared with Vanilla MoE, performance is comparable on Centers A and B but significantly improves on Center C, where the data distribution differs most, confirming that the center-specific router enhances adaptability under heterogeneous multicenter distributions.

We further analyzed the impact of varying the top-$k$ experts (Table~\ref{tab_ablation}(b)) and the total number of experts ($n$) within the proposed MoME framework (Table~\ref{tab_ablation}(c)). Using top-2 experts achieved balanced performance across centers, suggesting that optimal performance requires balancing the number of experts in relation to the number of centers. For the total number ($n$) of experts, decreasing $n$ degraded Center C performance while increasing $n$ degraded Center B performance, indicating that maintaining sufficient overlap in router selection enhances cross-center synergy. {We further examined two additional architectural components: the expert module design (Table~\ref{tab_ablation}(d)) and the choice of LLM (Table~\ref{tab_ablation}(e)). Compared with a Linear projection and a Modern, SiLU-activated expert, our ReLU-based MLP expert offers the best balance between expressiveness, robustness across centers, and parameter efficiency. For the LLM, LLaMA3-Instruct-8B (ours) provides the most consistent performance across centers compared with CLIP ViT-B \cite{radford2021clip} and Qwen3-8B \cite{yang2025qwen3}.} 

Next, Table~\ref{tab_ablation}(f) shows the impact of number of involved training centers on 0-shot generalization on closed centers. {The relative contribution of each training center to the closed-center generalization depends on its distributional similarity to the target closed center, which is consistent with the motivation of our MoME framework (Fig.~\ref{fig_intro}(a)). Specifically, training with Center A alone or adding few-shot samples from Center B, whose distribution differs substantially from the closed Centers D and E, provides only marginal or occasionally negative effect. In contrast, incorporating few-shot samples from Center C, which shares closer distributional characteristics with the closed centers, yields clear performance gains. }

\add{The computational cost of the MoME framework is further analyzed using the single NVIDIA RTX A6000 48GB GPU in Table~\ref{tab_cost}. Compared with the vision-only AI, multimodal approaches naturally require more parameters and operations due to the integration of LLM. Nevertheless, our MoME framework maintains a comparable parameter size and computational overhead relative to the multimodal AI baselines, yet achieves clear performance gains. This demonstrates that the proposed design improves performance while maintaining comparable computational cost, with only a modest increase in inference latency.} 
\add{In addition, computing multimodal AIs inevitably requires EMR curation as preprocessing through a local LLM, which may present a practical burden. However, with the rapid emergence of lightweight LLMs, we expect that multimodal AI will soon be readily accessible to clinical centers with modest infrastructure.}

\subsection{\add{MoME generalizability to unimodal tasks}}

\add{To evaluate the generalizability of the MoME module across different cancer types and radiotherapy tasks, we further conducted experiments on nasopharyngeal cancer using six CTV labels from the publicly available SegRap2025 challenge dataset \cite{segrap2025}. The dataset details and split strategy are provided in Supplementary Table V. As this public dataset is unimodal and lacks textual information, the interactive alignment module was excluded from the MoME framework. Supplementary Table VI shows that MoME consistently outperforms both single-center and multicenter baselines, demonstrating its potential for diverse radiotherapy delineation tasks. Nonetheless, broader validation across diverse types of cancer and therapeutic works remains a subject of future work.}

\section{Discussion and Conclusion}

Mixture of Multicenter Expert (MoME) framework is designed to tackle biased inference in medical AI by creating tailored center-specific paths that utilize small, diverse samples to address inter-institutional variability. The superiority of our MoME training is demonstrated by the ability of center-specific routers to enable the model to closely adapt to each center's treatment patterns.
Our method proves highly adaptable in clinical settings with restricted data sharing but necessary adaptation to new data distributions. 
Furthermore, few-shot fine-tuning using the selected center-specific router network with MoME outperforms traditional AI training mechanisms for optimizing pre-trained models in clinical deployment. This approach is particularly valuable for real-world applications with limited sample datasets.

{Despite the advantages, our multicenter MoME framework has several limitations that motivate future work:}

{\textit{(i) Reliance on small few-shot subsets at new centers.}
A primary limitation is the dependence on limited labeled samples for fine-tuning, which can restrict robustness when the cross-center distribution gap is large; for example, the absolute Dice on Closed Centers D and E remained below 0.7 despite MoME consistently outperforming multimodal AI baselines. However, our analysis of joint training across Centers A--C revealed clear synergistic gains when the added center exhibits complementary distributional characteristics, indicating that \textit{which} few-shot samples are selected matters more than \textit{how many}. Future work will therefore focus on distribution-aware few-shot sample selection that prioritizes underrepresented patterns, preserving the data-efficient nature of MoME while improving closed-center generalization.}

{\textit{(ii) One-step PTV delineation as a design choice.}
We adopted a one-step PTV delineation rather than the clinical practice-inspired step-wise (prostate $\rightarrow$ PTV) workflow, based on empirical and practical considerations supported by Supplementary Table~VII. Empirically, step-wise pipelines using either our in-house 3D U-Net or the public TotalSegmentator as the intermediate prostate module consistently underperform on the closed-center dataset, as suboptimal prostate masks propagate localization errors into the downstream PTV prediction rather than serving as a useful anatomical prior. While the current design does not fully mirror the clinical practice, we plan to revisit a controlled step-wise approach in follow-up work, leveraging a high-fidelity multimodal prostate segmentation module (e.g., integrating MR for accurate prostate contouring) as a learned prior to enable more accurate, evidence-based target delineation.}

{\textit{(iii) Deployment considerations and inter-clinician variability.}
For clinical deployment, MoME should accommodate not only inter-institutional but also inter-clinician variability, since individual clinicians' contouring strategies often differ even within the same center\cite{fotina2012critical, vinod2016review, caravatta2014inter}. A crucial next step is to extend the center-specific routing to clinician-wise distributions, treating each clinician as a fine-grained sub-population within a center. We envision an interactive correction system embedded in clinical practice, where AI-generated contours are iteratively refined through minimal, clinician-specific parameter updates along the existing MoME routing paths, enabling continual learning that reflects evolving treatment philosophies. This design would align the model with diverse clinical perspectives, support personalized patient care, and improve the framework's generalizability in real-world deployment.}

Building on these directions, our study marks a significant step toward collaboration across multicenter datasets despite the practical constraints of large-scale data collection, positioning the multimodal MoME as a promising candidate for complex and often debated clinical decision-making tasks where collaborative synergy and alignment with unique institutional strategies are essential.

\bibliographystyle{IEEEtran}
\bibliography{refs}

@article{burnet2004defining,
  title={Defining the tumour and target volumes for radiotherapy},
  author={Burnet, Neil G and Thomas, Simon J and Burton, Kate E and Jefferies, Sarah J},
  journal={Cancer Imaging},
  volume={4},
  number={2},
  pages={153},
  year={2004},
  publisher={BMC}
}

@article{tang2019clinically,
  title={Clinically applicable deep learning framework for organs at risk delineation in CT images},
  author={Tang, Hao and Chen, Xuming and Liu, Yang and Lu, Zhipeng and You, Junhua and Yang, Mingzhou and Yao, Shengyu and Zhao, Guoqi and Xu, Yi and Chen, Tingfeng and others},
  journal={Nature Machine Intelligence},
  volume={1},
  number={10},
  pages={480--491},
  year={2019},
  publisher={Nature Publishing Group UK London}
}

@article{lin2021deep,
  title={Deep learning for automatic target volume segmentation in radiation therapy: a review},
  author={Lin, Hui and Xiao, Haonan and Dong, Lei and Teo, Kevin Boon-Keng and Zou, Wei and Cai, Jing and Li, Taoran},
  journal={Quantitative Imaging in Medicine and Surgery},
  volume={11},
  number={12},
  pages={4847},
  year={2021},
  publisher={AME Publications}
}

@article{fotina2012critical,
  title={Critical discussion of evaluation parameters for inter-observer variability in target definition for radiation therapy},
  author={Fotina, I and L{\"u}tgendorf-Caucig, C and Stock, M and P{\"o}tter, R and Georg, D},
  journal={Strahlentherapie und Onkologie},
  volume={188},
  number={2},
  pages={160},
  year={2012},
  publisher={Springer Nature BV}
}

@article{vinod2016review,
  title={A review of interventions to reduce inter-observer variability in volume delineation in radiation oncology},
  author={Vinod, Shalini K and Min, Myo and Jameson, Michael G and Holloway, Lois C},
  journal={Journal of medical imaging and radiation oncology},
  volume={60},
  number={3},
  pages={393--406},
  year={2016},
  publisher={Wiley Online Library}
}

@article{caravatta2014inter,
  title={Inter-observer variability of clinical target volume delineation in radiotherapy treatment of pancreatic cancer: a multi-institutional contouring experience},
  author={Caravatta, Luciana and Macchia, Gabriella and Mattiucci, Gian Carlo and Sainato, Aldo and Cernusco, Nunzia LV and Mantello, Giovanna and Di Tommaso, Monica and Trignani, Marianna and De Paoli, Antonino and Boz, Gianni and others},
  journal={Radiation oncology},
  volume={9},
  pages={1--9},
  year={2014},
  publisher={Springer}
}

@article{barkati2016magnetic,
  title={Magnetic resonance imaging for prostate bed radiotherapy planning: an inter-and intra-observer variability study},
  author={Barkati, Maroie and Simard, Dany and Taussky, Daniel and Delouya, Guila},
  journal={Journal of Medical Imaging and Radiation Oncology},
  volume={60},
  number={2},
  pages={255--259},
  year={2016},
  publisher={Wiley Online Library}
}

@article{valicenti1999variation,
  title={Variation of clinical target volume definition in three-dimensional conformal radiation therapy for prostate cancer},
  author={Valicenti, Richard K and Sweet, John W and Hauck, Walter W and Hudes, Richard S and Lee, Tony and Dicker, Adam P and Waterman, Frank M and Anne, Pramila R and Corn, Benjamin W and Galvin, James M},
  journal={International Journal of Radiation Oncology* Biology* Physics},
  volume={44},
  number={4},
  pages={931--935},
  year={1999},
  publisher={Elsevier}
}

@manual{NCCN2024Prostate,
  title = {{NCCN Clinical Practice Guidelines in Oncology: Prostate Cancer (Version 4.2024)}},
  author = {{National Comprehensive Cancer Network}},
  year = {2024},
  note = {\url{https://www.nccn.org/professionals/physician_gls/pdf/prostate.pdf}},
}

@article{salembier2018estro,
  title={ESTRO ACROP consensus guideline on CT-and MRI-based target volume delineation for primary radiation therapy of localized prostate cancer},
  author={Salembier, Carl and Villeirs, Geert and De Bari, Berardino and Hoskin, Peter and Pieters, Bradley R and Van Vulpen, Marco and Khoo, Vincent and Henry, Ann and Bossi, Alberto and De Meerleer, Gert and others},
  journal={Radiotherapy and Oncology},
  volume={127},
  number={1},
  pages={49--61},
  year={2018},
  publisher={Elsevier}
}

@article{dal2023estro,
  title={ESTRO ACROP guideline on prostate bed delineation for postoperative radiotherapy in prostate cancer},
  author={Dal Pra, Alan and Dirix, Piet and Khoo, Vincent and Carrie, Christian and Cozzarini, Cesare and Fonteyne, Val{\'e}rie and Ghadjar, Pirus and Gomez-Iturriaga, Alfonso and Panebianco, Valeria and Zapatero, Almudena and others},
  journal={Clinical and translational radiation oncology},
  volume={41},
  pages={100638},
  year={2023},
  publisher={Elsevier}
}

@misc{liu2023artificial,
      title={Artificial General Intelligence for Radiation Oncology}, 
      author={Chenbin Liu and Zhengliang Liu and Jason Holmes and Lu Zhang and Lian Zhang and Yuzhen Ding and Peng Shu and Zihao Wu and Haixing Dai and Yiwei Li and Dinggang Shen and Ninghao Liu and Quanzheng Li and Xiang Li and Dajiang Zhu and Tianming Liu and Wei Liu},
      year={2023},
      eprint={2309.02590},
      archivePrefix={arXiv},
      primaryClass={physics.med-ph}
}

@article{huynh2020artificial,
  title={Artificial intelligence in radiation oncology},
  author={Huynh, Elizabeth and Hosny, Ahmed and Guthier, Christian and Bitterman, Danielle S and Petit, Steven F and Haas-Kogan, Daphne A and Kann, Benjamin and Aerts, Hugo JWL and Mak, Raymond H},
  journal={Nature Reviews Clinical Oncology},
  volume={17},
  number={12},
  pages={771--781},
  year={2020},
  publisher={Nature Publishing Group UK London}
}

@article{kirillov2023segment,
  title={Segment anything},
  author={Kirillov, Alexander and Mintun, Eric and Ravi, Nikhila and Mao, Hanzi and Rolland, Chloe and Gustafson, Laura and Xiao, Tete and Whitehead, Spencer and Berg, Alexander C and Lo, Wan-Yen and others},
  journal={arXiv preprint arXiv:2304.02643},
  year={2023}
}

@article{zhang2023segment,
  title={Segment Anything Model (SAM) for Radiation Oncology},
  author={Zhang, Lian and Liu, Zhengliang and Zhang, Lu and Wu, Zihao and Yu, Xiaowei and Holmes, Jason and Feng, Hongying and Dai, Haixing and Li, Xiang and Li, Quanzheng and others},
  journal={arXiv preprint arXiv:2306.11730},
  year={2023}
}

@inproceedings{cciccek20163d,
  title={3D U-Net: learning dense volumetric segmentation from sparse annotation},
  author={{\c{C}}i{\c{c}}ek, {\"O}zg{\"u}n and Abdulkadir, Ahmed and Lienkamp, Soeren S and Brox, Thomas and Ronneberger, Olaf},
  booktitle={Medical Image Computing and Computer-Assisted Intervention--MICCAI 2016: 19th International Conference, Athens, Greece, October 17-21, 2016, Proceedings, Part II 19},
  pages={424--432},
  year={2016},
  organization={Springer}
}

@inproceedings{radford2021clip,
  title={Learning transferable visual models from natural language supervision},
  author={Radford, Alec and Kim, Jong Wook and Hallacy, Chris and Ramesh, Aditya and Goh, Gabriel and Agarwal, Sandhini and Sastry, Girish and Askell, Amanda and Mishkin, Pamela and Clark, Jack and others},
  booktitle={International Conference on Machine Learning},
  pages={8748--8763},
  year={2021},
  organization={PMLR}
}

@article{crum2006generalized,
  title={Generalized overlap measures for evaluation and validation in medical image analysis},
  author={Crum, William R and Camara, Oscar and Hill, Derek LG},
  journal={IEEE transactions on medical imaging},
  volume={25},
  number={11},
  pages={1451--1461},
  year={2006},
  publisher={IEEE}
}

@article{shi2022deep,
  title={Deep learning empowered volume delineation of whole-body organs-at-risk for accelerated radiotherapy},
  author={Shi, Feng and Hu, Weigang and Wu, Jiaojiao and Han, Miaofei and Wang, Jiazhou and Zhang, Wei and Zhou, Qing and Zhou, Jingjie and Wei, Ying and Shao, Ying and others},
  journal={Nature Communications},
  volume={13},
  number={1},
  pages={6566},
  year={2022},
  publisher={Nature Publishing Group UK London}
}

@article{harrison2022machine,
  title={Machine learning for auto-segmentation in radiotherapy planning},
  author={Harrison, K and Pullen, H and Welsh, C and Oktay, O and Alvarez-Valle, J and Jena, R},
  journal={Clinical Oncology},
  volume={34},
  number={2},
  pages={74--88},
  year={2022},
  publisher={Elsevier}
}

@article{oh2024llm,
  author  = {Yujin Oh and Sangjoon Park and Hwa Kyung Byun and Yeona Cho and Ik Jae Lee and Jin Sung Kim and Jong Chul Ye},
  title   = {LLM-driven multimodal target volume contouring in radiation oncology},
  journal = {Nature Communications},
  year    = {2024},
  volume  = {15},
  number  = {1},
  pages   = {9186},
  doi     = {10.1038/s41467-024-53387-y},
  issn    = {2041-1723}
}

@article{shazeer2017outrageously,
  title={Outrageously large neural networks: The sparsely-gated mixture-of-experts layer},
  author={Shazeer, Noam and Mirhoseini, Azalia and Maziarz, Krzysztof and Davis, Andy and Le, Quoc and Hinton, Geoffrey and Dean, Jeff},
  journal={arXiv preprint arXiv:1701.06538},
  year={2017}
}

@article{rajendran2024auto,
  title={Auto-delineation of treatment target volume for radiation therapy using large language model-aided multimodal learning},
  author={Rajendran, Praveenbalaji and Chen, Yizheng and Qiu, Liang and Niedermayr, Thomas and Liu, Wu and Buyyounouski, Mark and Bagshaw, Hilary and Han, Bin and Yang, Yong and Kovalchuk, Nataliya and others},
  journal={International Journal of Radiation Oncology* Biology* Physics},
  year={2024},
  publisher={Elsevier}
}

@inproceedings{yu2024boosting,
  title={Boosting continual learning of vision-language models via mixture-of-experts adapters},
  author={Yu, Jiazuo and Zhuge, Yunzhi and Zhang, Lu and Hu, Ping and Wang, Dong and Lu, Huchuan and He, You},
  booktitle={Proceedings of the IEEE/CVF Conference on Computer Vision and Pattern Recognition},
  pages={23219--23230},
  year={2024}
}

@article{van2020brain,
  title={Brain-inspired replay for continual learning with artificial neural networks},
  author={Van de Ven, Gido M and Siegelmann, Hava T and Tolias, Andreas S},
  journal={Nature communications},
  volume={11},
  number={1},
  pages={4069},
  year={2020},
  publisher={Nature Publishing Group UK London}
}

@article{rypesc2024divide,
  title={Divide and not forget: Ensemble of selectively trained experts in Continual Learning},
  author={Rype{\'s}{\'c}, Grzegorz and Cygert, Sebastian and Khan, Valeriya and Trzci{\'n}ski, Tomasz and Zieli{\'n}ski, Bartosz and Twardowski, Bart{\l}omiej},
  journal={arXiv preprint arXiv:2401.10191},
  year={2024}
}

@article{zhang2024generalist,
  title={A generalist vision--language foundation model for diverse biomedical tasks},
  author={Zhang, Kai and Zhou, Rong and Adhikarla, Eashan and Yan, Zhiling and Liu, Yixin and Yu, Jun and Liu, Zhengliang and Chen, Xun and Davison, Brian D and Ren, Hui and others},
  journal={Nature Medicine},
  pages={1--13},
  year={2024},
  publisher={Nature Publishing Group US New York}
}

@misc{tianfairseg,
      title={FairSeg: A Large-Scale Medical Image Segmentation Dataset for Fairness Learning Using Segment Anything Model with Fair Error-Bound Scaling}, 
      author={Yu Tian and Min Shi and Yan Luo and Ava Kouhana and Tobias Elze and Mengyu Wang},
      year={2024},
      eprint={2311.02189},
      archivePrefix={arXiv},
      primaryClass={cs.CV},
      url={https://arxiv.org/abs/2311.02189}, 
      booktitle={The Twelfth International Conference on Learning Representations}
}

@article{ktena2024generative,
  title={Generative models improve fairness of medical classifiers under distribution shifts},
  author={Ktena, Ira and Wiles, Olivia and Albuquerque, Isabela and Rebuffi, Sylvestre-Alvise and Tanno, Ryutaro and Roy, Abhijit Guha and Azizi, Shekoofeh and Belgrave, Danielle and Kohli, Pushmeet and Cemgil, Taylan and others},
  journal={Nature Medicine},
  pages={1--8},
  year={2024},
  publisher={Nature Publishing Group US New York}
}

@inproceedings{
oh2025dmoe,
title={Distribution-aware Fairness Learning in Medical Image Segmentation From A Control-Theoretic Perspective},
author={Yujin Oh and Pengfei Jin and Sangjoon Park and Sekeun Kim and Siyeop yoon and Jin Sung Kim and Kyungsang Kim and Xiang Li and Quanzheng Li},
booktitle={Forty-second International Conference on Machine Learning},
year={2025},
url={https://openreview.net/forum?id=BUONdewsBa}
}

@article{segrap2025,
	author = {Luo, Xiangde and Liao, Wenjun and Zhao, Yue and Qiu, Youjing and Xu, Jinfeng and He, Yuan and Huang, Hui and Li, Lu and Zhang, Shichuan and Fu, Jia and Wang, Guotai and Zhang, Shaoting},
	date = {2024/10/04},
	doi = {10.1038/s41597-024-03890-0},
	id = {Luo2024},
	isbn = {2052-4463},
	journal = {Scientific Data},
	number = {1},
	pages = {1085},
	title = {A multicenter dataset for lymph node clinical target volume delineation of nasopharyngeal carcinoma},
	url = {https://doi.org/10.1038/s41597-024-03890-0},
	volume = {11},
	year = {2024}}

@article{doi:10.1148/ryai.230024,
author = {Wasserthal, Jakob and Breit, Hanns-Christian and Meyer, Manfred T. and Pradella, Maurice and Hinck, Daniel and Sauter, Alexander W. and Heye, Tobias and Boll, Daniel T. and Cyriac, Joshy and Yang, Shan and Bach, Michael and Segeroth, Martin},
title = {TotalSegmentator: Robust Segmentation of 104 Anatomic Structures in CT Images},
journal = {Radiology: Artificial Intelligence},
volume = {5},
number = {5},
pages = {e230024},
year = {2023},
doi = {10.1148/ryai.230024},
URL = { 
    
        https://doi.org/10.1148/ryai.230024

},
eprint = { 
    
        https://doi.org/10.1148/ryai.230024
},
}

@inproceedings{jeong2026multiexpert,
    title={Multi-Expert Distributionally Robust Optimization for Out-of-Distribution Generalization},
    author={Jinyong Jeong and Hyungu Kahng and Seoung Bum Kim},
    booktitle={The Thirty-ninth Annual Conference on Neural Information Processing Systems},
    year={2026},
    url={https://openreview.net/forum?id=Lz5BUjArK4}
}

@misc{panaganti2026groupd,
      title={Group Distributionally Robust Optimization-Driven Reinforcement Learning for LLM Reasoning}, 
      author={Kishan Panaganti and Zhenwen Liang and Wenhao Yu and Haitao Mi and Dong Yu},
      year={2026},
      eprint={2601.19280},
      archivePrefix={arXiv},
      primaryClass={cs.LG},
      url={https://arxiv.org/abs/2601.19280}, 
}

@misc{mcmahan2023fedavg,
      title={Communication-Efficient Learning of Deep Networks from Decentralized Data}, 
      author={H. Brendan McMahan and Eider Moore and Daniel Ramage and Seth Hampson and Blaise Agüera y Arcas},
      year={2023},
      eprint={1602.05629},
      archivePrefix={arXiv},
      primaryClass={cs.LG},
      url={https://arxiv.org/abs/1602.05629}, 
}

@misc{li2020fedprox,
      title={Federated Optimization in Heterogeneous Networks}, 
      author={Tian Li and Anit Kumar Sahu and Manzil Zaheer and Maziar Sanjabi and Ameet Talwalkar and Virginia Smith},
      year={2020},
      eprint={1812.06127},
      archivePrefix={arXiv},
      primaryClass={cs.LG},
      url={https://arxiv.org/abs/1812.06127}, 
}

@article{yang2025qwen3,
  title={Qwen3 technical report},
  author={Yang, An and Li, Anfeng and Yang, Baosong and Zhang, Beichen and Hui, Binyuan and Zheng, Bo and Yu, Bowen and Gao, Chang and Huang, Chengen and Lv, Chenxu and others},
  journal={arXiv preprint arXiv:2505.09388},
  year={2025}
}


\onecolumn
\section*{Biography Section}
\begin{multicols}{2}

\begin{IEEEbiography}[{\includegraphics[width=0.9in,clip,keepaspectratio]{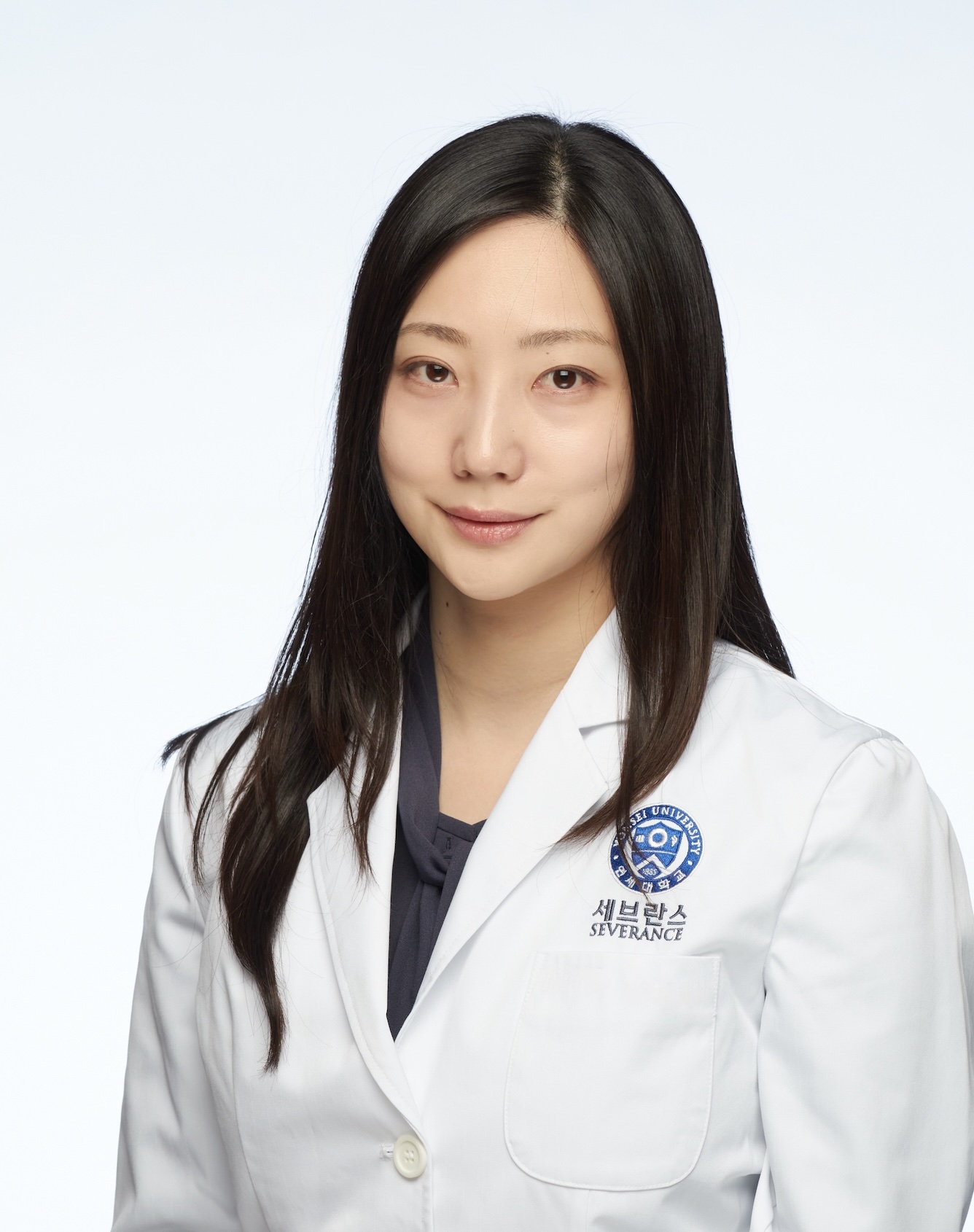}}]{Yujin Oh} is an Assistant Professor in the Department of Biomedical Systems Informatics, College of Medicine, Yonsei University. She received her Ph.D. from the Graduate School of AI, Korea Advanced Institute of Science and Technology (KAIST), and was previously a Postdoctoral Researcher at Massachusetts General Hospital (MGH) and Harvard Medical School (HMS). Her research lies at the intersection of multimodal data, engineering theory, and clinical practice, building systems that are accurate, interpretable, and deployable in healthcare.
\end{IEEEbiography}

\vspace{-30pt}
\begin{IEEEbiography}[{\includegraphics[width=0.9in,clip,keepaspectratio]{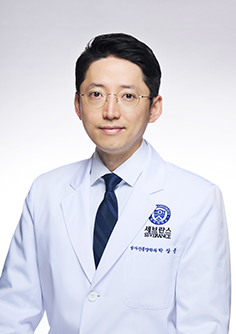}}]{Sangjoon Park}
is an Assistant Professor of Radiation Oncology at Yonsei University College of Medicine. He completed his residency at Yonsei Cancer Center and earned a Ph.D. in engineering from Korea Advanced Institute of Science and Technology (KAIST). His research focuses on LLM, foundation models, and multimodal AI in radiation oncology.
\end{IEEEbiography}

\vspace{-30pt}
\begin{IEEEbiography}[{\includegraphics[width=0.9in, clip,keepaspectratio]{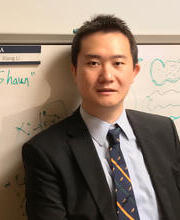}}]{Xiang Li} is an Assistant Professor at the Mass General Hospital (MGH) and Harvard Medical School (HMS). He received his Ph.D. degree from the Department of Computer Science at the University of Georgia. His research focuses on the Artificial General Intelligence (AGI) to tackle the practical challenges of applying AI in a complex clinical context.
\end{IEEEbiography}

\vspace{-30pt}
\begin{IEEEbiographynophoto}{Pengfei Jin} is a Postdoctoral Researcher at the MGH and HMS. He received his Ph.D. degree from the Department of Mathematics at Peking University. 
\end{IEEEbiographynophoto}

 \vspace{-30pt}
\begin{IEEEbiographynophoto}{Yi Wang} is an Assistant Professor of Radiation Oncology at MGH and HMS. He received his Ph.D. in Biomedical Engineering from the University of Michigan, and residency training at HMS. 
\end{IEEEbiographynophoto}

\vspace{-30pt}
\begin{IEEEbiographynophoto}{Jonathan Paly} is an Assistant Radiation Oncologist at MGH and HMS. His research focuses on integrating cutting-edge AI technology into radiation oncology to address critical clinical needs.
\end{IEEEbiographynophoto}

\vspace{-30pt}
\begin{IEEEbiographynophoto}{Jason Efstathiou} is a Professor of Radiation Oncology at MGH and HMS. As a Radiation Oncologist, he is actively involved in translational science, evaluating biomarkers for prostate and bladder cancer outcomes.  
\end{IEEEbiographynophoto}

\vspace{-30pt}
\begin{IEEEbiographynophoto}{Annie Chan} is an Associate Professor and Director of Head and Neck in the Department of Radiation Oncology at MGH and HMS. Her interests include AI-guided clinical target volume delineation and clinical evaluation.
\end{IEEEbiographynophoto}

\vspace{-30pt}
\begin{IEEEbiographynophoto}{Jun Won Kim}
 is a Professor and Department Chair of Radiation Oncology at Gangnam Severance Hospital, Yonsei University College of Medicine. His research focuses on prostate cancer and MR-Linac-based radiotherapy.
\end{IEEEbiographynophoto}

\vspace{-30pt}
\begin{IEEEbiographynophoto}{Hwa Kyung Byun}
 is an Assistant Professor of Radiation Oncology at Yonsei University College of Medicine. Her research focuses on breast and liver cancers, as well as large language models in radiation oncology.
\end{IEEEbiographynophoto}

\vspace{-30pt}
\begin{IEEEbiographynophoto}{Ik Jae Lee}
is a Professor of Radiation Oncology at Yonsei University College of Medicine. His research focuses on prostate, liver, and breast cancers.
\end{IEEEbiographynophoto}

\vspace{-30pt}
\begin{IEEEbiographynophoto}{Jaeho Cho}
 is a Professor of Radiation Oncology at Yonsei University College of Medicine. His research focuses on prostate and lung cancers.
\end{IEEEbiographynophoto}

\vspace{-30pt}
\begin{IEEEbiographynophoto}{Chan Woo Wee}
is an Assistant Professor of Radiation Oncology at Yonsei University College of Medicine. His research focuses on the clinical and genetic aspects of brain tumors and prostate cancer radiotherapy.
\end{IEEEbiographynophoto}

\vspace{-30pt}
\begin{IEEEbiographynophoto}{Peng Shu} is a Ph.D. student of Computer Science in University of Georgia (UGA). He received his master degree in the University of Edinburgh, UK and became a full-time SW engineer at Hisilicon. 
\end{IEEEbiographynophoto}

\vspace{-30pt}
\begin{IEEEbiographynophoto}{Peilong Wang} is a Research Fellow of Radiation Oncology at Mayo Clinic Arizona. He received his Ph.D. in Physics from Southern Methodist University. His research focuses on medical physics, AI, and medical imaging.
\end{IEEEbiographynophoto}

 \columnbreak

\vspace{-35pt}
\begin{IEEEbiographynophoto}{Caiwen Jiang}
is currently a Postdoctoral Fellow at Mayo Clinic Arizona. He received his Ph.D. degree in Computer Science from ShanghaiTech University, Shanghai, China. His research interests include medical image processing and the application of AI in radiation oncology. 
\end{IEEEbiographynophoto}

\vspace{-30pt}
\begin{IEEEbiographynophoto}{Nathan Yu} is an Assistant Professor of Radiation Oncology at Mayo Clinic Arizona. He received his M.D. from  University of California, San Diego. 
\end{IEEEbiographynophoto}

\vspace{-30pt}
\begin{IEEEbiographynophoto}{Jason Holmes} is a Researcher of Radiation Oncology at Mayo Clinic Arizona. He received a Ph.D. in physics from Arizona State University on the topic of radiation detection and imaging and subsequently became a research fellow.
\end{IEEEbiographynophoto}

\vspace{-30pt}
\begin{IEEEbiographynophoto}{Jong Chul Ye (Fellow, IEEE)} is a Professor of the Graduate School of AI at KAIST. He received his Ph.D. from Purdue University, West Lafayette. He is currently an Associate Editor for IEEE Transactions on Medical Imaging, a Senior Editor of IEEE Signal Processing Magazine, and an Executive Editor of Biological Imaging. His research interest is in machine learning applications and theory for biomedical imaging and computer vision.
\end{IEEEbiographynophoto}

\vspace{-30pt}
\begin{IEEEbiography}[{\includegraphics[width=0.9in,clip,keepaspectratio]{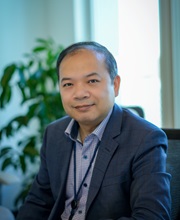}}]{Quanzheng Li} is an Associate Professor at Mass General Hospital (MGH) and Harvard Medical School (HMS). His research interests include deep learning on multimodality clinical data, including imaging and electronic health records, for screening, risk prediction, diagnosis, treatment optimization, and prognosis of various diseases.
\end{IEEEbiography}

\vspace{-30pt}
\begin{IEEEbiography}[{\includegraphics[width=0.9in,clip,keepaspectratio]{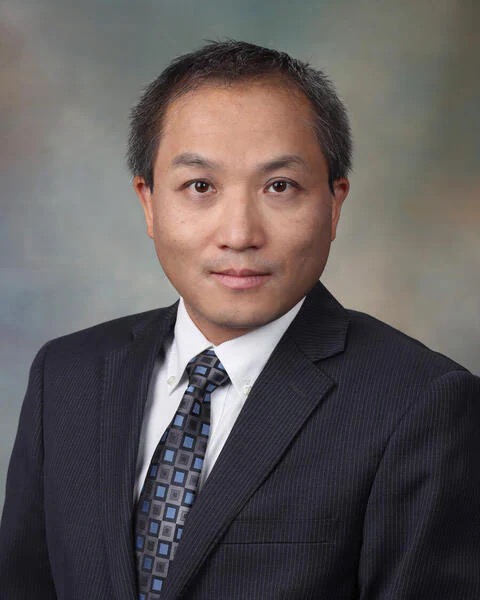}}]{Wei Liu} is a Professor of Radiation Oncology and Vice Department Chair of Physics Research of Radiation Oncology of Mayo Clinic Arizona. He received Ph.D. from Princeton University. He is currently an Associate Editor for IEEE Transactions on Medical Imaging, International Journal of Radiation Oncology • Biology • Physics, and Medical Physics and an Editorial Board member of Physics in Medicine and Biology and Radiotherapy and Oncology. 
He is a Fellow of American Association of Physicists in Medicine (AAPM). He is also a Fellow of American Institute for Medical Biological Engineering (AIMBE) and currently serves as Co-Lead of Workstream in Data Science, AI, and LLMs of Mayo Clinic Comprehensive Cancer Center in Arizona. 
\end{IEEEbiography}

\vspace{-25pt}
\begin{IEEEbiography}[{\includegraphics[width=0.9in,clip,keepaspectratio]{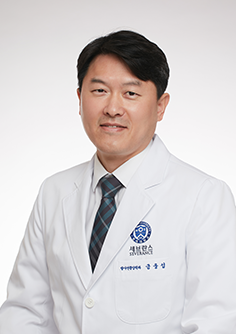}}]{Woong Sub Koom}
is a Professor and Department Chair of Radiation Oncology at Yonsei University College of Medicine. His research focuses on carbon ion therapy and genitourinary cancers, including prostate cancer.
\end{IEEEbiography}

\vspace{-30pt}
\begin{IEEEbiography}[{\includegraphics[width=0.9in,clip,keepaspectratio]{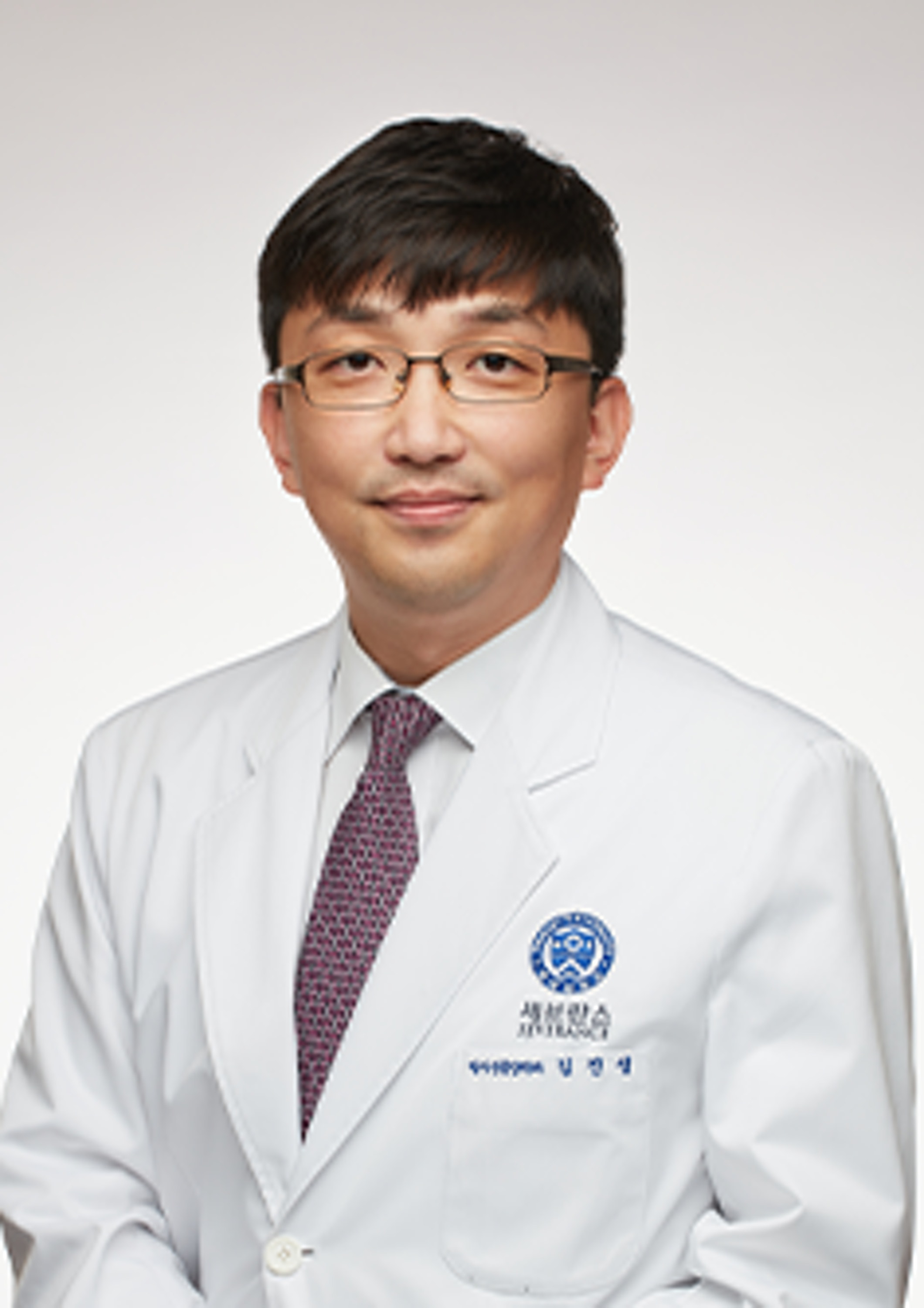}}]{Jin Sung Kim}
is a Professor at Yonsei University College of Medicine, specializing in medical physics and carbon ion therapy. His research focuses on advanced radiation therapy, AI-driven segmentation, and LLM-based multimodal image analysis for oncology decision support.
\end{IEEEbiography}

\vspace{-30pt}
\begin{IEEEbiography}[{\includegraphics[width=0.9in,clip,keepaspectratio]{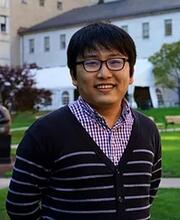}}]{Kyungsang Kim} is an Assistant Professor at Mass General Hospital (MGH) and Harvard Medical School (HMS). He received his Ph.D. from the Department of Bio and Brain Engineering at KAIST. His research focuses on medical AI and signal processing, leveraging multicenter and multimodal imaging with clinical information to address AI bias.
\end{IEEEbiography}
\end{multicols}

\newpage



\clearpage
\onecolumn

 \section*{\huge Supplementary Material}

 \vspace{0.75cm}

 \setcounter{page}{1}
 \setcounter{figure}{0}
 \setcounter{table}{0}

 \renewcommand{\figurename}{Supplementary Figure}
 \renewcommand{\tablename}{Supplementary Table}
 \setcounter{section}{0} 
 \renewcommand{\thesection}{Supplementary Section \roman{section}} 

 \section{Prostate cancer data characteristics} 


 We utilized datasets from five different centers, which is provided in Supplementary Table I. For model training, we utilized the largest dataset from Center A (Yonsei Cancer Center, Seoul, South Korea). A total of 943 primary prostate cancer patients were randomly split, with 774 patients used for training and 169 for internal validation. Center B (Yongin Severance Hospital, Yongin, South Korea) contributed data from 137 patients. For fine-tuning, 10, 15, or 20 patients were used under different experimental conditions, with the remaining 117 patients reserved for external validation.
 Similarly, Center C (Gangnam Severance Hospital, Seoul, South Korea) provided data from 149 patients. We used 10, 15, or 20 patients for fine-tuning, while the remaining 129 were used for external validation.
 For Center D (MGH, Boston, MA, USA), a total of 67 patients were collected, with 10, 15, or 20 patients used for fine-tuning in the closed center setting, and the remaining 47 used for external validation.
 Finally, Center E (Mayo Clinic, Phoenix, AZ, USA) contributed data from 110 patients, with 10, 15, or 20 patients used for fine-tuning in the closed center environment, and the remaining 90 patients utilized for external validation. The data collected for this study were ethically approved by the Institutional Review Boards (IRB) of the Department of Radiation Oncology at Yonsei Cancer Center, Department of Radiation Oncology at Yongin Severance Hospital, and Department of Radiation Oncology at Gangnam Severance Hospital (IRB numbers 4-2023-0179, 9-2023-0161, and 3-2023-0396, respectively), Department of Radiation Oncology at Mayo Clinic (IRB number 13-005709), and Massachusetts General Hospital (IRB number 2021P002249). The requirement for informed consent was waived due to the retrospective nature of the study.

 As illustrated in Fig. 1(a), Centers A, B, and C are located in South Korea, and while the patient characteristics vary based on the size and location of the centers, they share similar ethnic backgrounds. In contrast, Centers D and E, located in the United States, have a more diverse racial composition compared to the Korean centers (A–C). To address the potential limitations of data sharing between countries, we simulated a closed center environment for Centers D and E. In this scenario, direct data sharing is restricted, and only model weights are transferred. This allowed us to evaluate the feasibility of fine-tuning the MoME model in an in-house setting without exchanging sensitive patient data. In terms of clinical characteristics, Center A had a higher proportion of locally advanced cases, with a higher tendency towards elevated T stages. In contrast, Centers B and C showed fewer cases with high T stages. This trend was even more pronounced in the U.S. centers (D and E), where T stages were generally even lower than those observed in Centers B and C. Across all institutions, N stage showed minimal variation, with most cases being node-negative, which provided an ideal setting to evaluate institutional policies regarding prophylactic nodal irradiation (PNI).
 Similar to the T stage trend, the Korean centers (A–C) generally had higher Gleason scores, indicating a greater prevalence of advanced tumors. This was also reflected in the initial PSA values (iPSA), where the Korean institutions reported higher values compared to the U.S. centers. Among them, Center A had the highest iPSA values overall, while the U.S. centers exhibited comparatively lower values.
 There were also notable differences in the rates of prostatectomy between the Korean and U.S. centers. In the Korean centers, 40\% to 80\% of patients underwent surgery, whereas approximately more than 70\% of patients in the U.S. centers received definitive radiotherapy without surgery.
 These differences in surgical rates influenced the treatment intent. In the Korean centers, around 50\% to over 80\% of patients received adjuvant or salvage radiotherapy after surgery, while in the U.S. centers, most patients received definitive radiotherapy without undergoing surgery.
 Regarding imaging acquisition settings, Centers A and B used similar devices and followed comparable protocols. While Centers C and E employed different settings from A and B, they were closely aligned with each other in their imaging acquisition approaches. In contrast, Center D utilized a distinct combination of devices and protocols, further differentiating it from the other centers.
 These similarities and differences in imaging acquisition settings, patient demographics (e.g., the similarity between Centers A and B), and clinical practices (e.g., the notable differences between the remaining centers) provided a structured environment to systematically evaluate the effectiveness of MoME in adapting the model to various national and institutional treatment strategies.


 \section{Implementation details} 

For data preprocessing, all chest CT images and PTV labels are resampled to a uniform voxel spacing of 1.0 × 1.0 × 3.0 mm$^{3}$. The image intensities are truncated between -200 and 250 Hounsfield units (HU) and linearly normalized to a range between 0 and 1. For preprocessing of EMR data, we utilize the Vicuna-33B\footnote{\url{https://huggingface.co/lmsys/vicuna-33b-v1.3}} checkpoint on a local server to curate clinical data, as summarized in Supplementary Table II. 

For multimodal radiotherapy target delineation, we employ a 3D Residual U-Net as an image module backbone and a pre-trained LLaMA3-8B-Instruct\footnote{\url{https://huggingface.co/meta-llama/Meta-Llama-3-8B-Instruct}} as a language module. 
During network training, 3D patches of 384 × 384 × 128 pixels are randomly cropped to include the entire pelvic region, along with the corresponding clinical data, using a batch size of 2. For evaluation, the full 3D CT volumes are processed with a sliding window approach, using the same patch size for training. Throughout training, the entire LLM module is kept frozen, while the image encoder/decoder modules, interactive alignment modules, and text prompts are optimized. The length of learnable text prompts $M$ is set to 32 and the total length of total clinical data $L$ is set to 96.  We set the hyperparameter top-$k$ as 2, and $n$ as 8. The loss function combine binary cross-entropy and Dice loss, with equal weights of 1.0. The network is optimized using the AdamW optimizer, with an initial learning rate of 0.0001, for 100 training epochs. For multicenter training, we utilize the entire Center A training dataset, combined with 1-shot, 2-shot, and 3-shot samples from Centers B and C for each trial with 1-shot validation samples. For fine-tuning the network, the learning rate is reduced to 0.00001, and the network parameters are optimized for up to 500 fine-tuning epochs. For fine-tuning the network to each closed center, we utilize 1-shot, 2-shot, and 3-shot samples with 1-shot validation samples. The few-shot samples are randomly selected based on 5 levels of PSA clusters, defined as: 0 ($\mathrm{PSA} < 5.0$), 1 ($5.0 \leq \mathrm{PSA} < 10.0$), 2 ($10.0 \leq \mathrm{PSA} < 20.0$), 3 ($20.0 \leq \mathrm{PSA} < 30.0$), and 4 ($\mathrm{PSA} \geq 30.0$).

The network is implemented using the open-source library MONAI\footnote{\url{https://github.com/project-monai/monai}}. All experiments are conducted using PyTorch in Python, leveraging CUDA 11.4 on a single NVIDIA RTX A6000 48GB GPU. For in-house model fine-tuning, we further utilize a single NVIDIA A100 80GB GPU. 

For statistical analysis, we employ the non-parametric bootstrap method to estimate confidence intervals (CIs) for each metric. We perform 1,000 resampling iterations with replacement from the original dataset to generate bootstrap samples. The mean values and 95\% CIs are then derived from the relative frequency distributions of these bootstrap samples. Statistical comparisons between groups are conducted using a two-tailed Student's paired t-test. The determination of the sample size is not based on statistical methods. No data are excluded from the analyzes, and the experiments were not randomized. The investigator was not blinded to the allocation during the experiments or outcome assessment.


 \begin{table*}[!h]
   \centering
   \caption{{Details of prostate cancer data partitioning and characteristics for each center.}}
   \resizebox{0.95\linewidth}{!}{
     \begin{tabular}{lcccccc}
     \toprule
    
     \multirow{1}{*}{\shortstack[c]{\bf{Center}}} & \multicolumn{2}{c}{\textbf{Center A}} & \multicolumn{1}{c}{\textbf{Center B}} & \multicolumn{1}{c}{{\bf{Center C}}} & \multicolumn{1}{c}{{\bf{Closed Center D}}} & \multicolumn{1}{c}{{\bf{Closed Center E}}}  \\
      \cmidrule(l){1-1} \cmidrule(l){2-3}  \cmidrule(l){4-4}  \cmidrule(l){5-5}  \cmidrule(l){6-6}  \cmidrule(l){7-7} 
     
      \textbf{{Hospital}} & \multicolumn{2}{c}{{Yonsei Cancer Center}}  & Yongin Severance & Gangnam Severance & MGH & MAYO Clinic  \\
      \cmidrule(l){1-1} \cmidrule(l){2-3}  \cmidrule(l){4-4}  \cmidrule(l){5-5}  \cmidrule(l){6-6}  \cmidrule(l){7-7} 
     
      \multirow{2}{*}{\shortstack[c]{\bf{Data split}}} & \multirow{2}{*}{{Train (n=774)}} 
      & \multirow{2}{*}{{Test (n=169)}}
      & \multicolumn{1}{c}{{Train (n=10/15/20$^{\dagger}$)}} 
      & \multicolumn{1}{c}{{{Train (n=10/15/20$^{\dagger}$)}}} 
      & \multicolumn{1}{c}{{{Fine-tune (n=10/15/20$^{\dagger}$)}}} 
      & \multicolumn{1}{c}{{{Fine-tune (n=10/15/20$^{\dagger}$)}}} \\

      &
      &  
      & \multicolumn{1}{c}{{Test (n=117)}} 
      & \multicolumn{1}{c}{{{Test (n=129)}}} 
      & \multicolumn{1}{c}{{{Test (n=47)}}}
      & \multicolumn{1}{c}{{{Test (n=90)}}} \\

     \cmidrule(l){1-1} \cmidrule(l){2-2} \cmidrule(l){3-3}  \cmidrule(l){4-4}  \cmidrule(l){5-5}  \cmidrule(l){6-6}  \cmidrule(l){7-7} 
    
     \shortstack[c]{\textbf{{Label Description}}\\{}} & \\
     {  }0: Background   &\\
     {  }1: PTV   &\\

     \cmidrule(l){1-1} \cmidrule(l){2-2} \cmidrule(l){3-3}  \cmidrule(l){4-4}  \cmidrule(l){5-5}  \cmidrule(l){6-6}  \cmidrule(l){7-7} 

     \textbf{{T stage}} & \\
     {  }T1   & 31 (4.1\%) & 5 (3.0\%) & 1 (0.7\%) & 10 (6.7\%)&  39 (58.2\%) & 38 (34.5\%) \\
     {  }T2   & 231 (30.6\%) & 58 (34.3\%) & 55 (40.1\%) & 78 (52.3\%) & 13 (19.4\%) & 32 (29.1\%) \\
     {  }T3   & 435 (57.7\%) & 100 (59.2\%) & 67 (48.9\%) & 49 (32.9\%) & 15 (22.4\%) & 36 (32.7\%) \\
     {  }T4   & 57 (7.6\%) & 6 (3.6\%) & 14 (10.2\%) & 12 (8.1\%) & 0 (0\%) & 4 (3.6\%) \\
     \cmidrule(l){1-1} \cmidrule(l){2-2} \cmidrule(l){3-3}  \cmidrule(l){4-4}  \cmidrule(l){5-5}  \cmidrule(l){6-6}  \cmidrule(l){7-7} 
    
     \textbf{{N stage}} & & \\
     {  }N0   & 676 (89.7\%) & 150 (89.9\%) & 118 (86.1\%) & 137 (91.9\%) & 57 (85.1\%) & 101 (91.8\%) \\
     {  }N1   & 78 (10.3\%) & 19 (10.1\%) & 19 (13.9\%) & 12 (8.1\%) & 10 (14.9\%) & 9 (8.2\%) \\
     \cmidrule(l){1-1} \cmidrule(l){2-2} \cmidrule(l){3-3}  \cmidrule(l){4-4}  \cmidrule(l){5-5}  \cmidrule(l){6-6}  \cmidrule(l){7-7} 
    
     \textbf{{Gleason score}} & \\
     {  }5 (2+3)   & 20 (2.6\%) & 0 (0\%) & 0 (0\%) & 0 (0\%) & 0 (0\%) & 3 (2.7\%) \\
     {  }6 (3+3)   & 42 (5.4\%) & 6 (3.2\%) & 17 (12.4\%) & 17 (11.4\%) & 6 (9.2\%) & 12 (10.9\%) \\
     {  }7 (3+4, 4+3)   & 318 (41.1\%) & 58 (36.0\%) & 57 (41.6\%) & 70 (47.0\%) & 38 (58.5\%) & 57 (51.8\%) \\
     {  }8 (3+5, 4+4, 5+3)  & 150 (19.4\%) & 43 (25.4\%) & 22 (16.1\%) & 22 (14.8\%) & 8 (12.3\%) & 17 (15.5\%) \\
     {  }9 (4+5, 5+4)   & 225 (29.1\%) & 58 (33.3\%) & 35 (25.5\%) & 35 (23.5\%) & 13 (20.0\%) & 19 (17.3\%) \\
     {  }10 (5+5)   & 19 (2.5\%) & 4 (2.1\%) & 6 (4.4\%) & 5 (3.4\%) & 0 (0\%) & 2 (1.8\%) \\
    
     \cmidrule(l){1-1} \cmidrule(l){2-2} \cmidrule(l){3-3}  \cmidrule(l){4-4}  \cmidrule(l){5-5}  \cmidrule(l){6-6}  \cmidrule(l){7-7} 
    
     \textbf{{Initial PSA}} & 39.3 (0.3-3865.0) & 39.3 (0.6-682.0) & 27.7 (0.9-217.0)  & 22.2 (2.99-281.67) & 12.5 (0.2-126.0) & 9.5 (0-156.0) \\
    
     \cmidrule(l){1-1} \cmidrule(l){2-2} \cmidrule(l){3-3}  \cmidrule(l){4-4}  \cmidrule(l){5-5}  \cmidrule(l){6-6}  \cmidrule(l){7-7} 
    
     \textbf{{Prostatectomy}} & \\
     {  }Yes   & 511 (66.0\%) & 129 (76.3\%) & 55 (40.1\%) & 70 (47.0\%) & 9 (13.4\%) & 35 (31.8\%) \\
     {  }No   & 263 (34.0\%) & 40 (23.7\%) & 82 (59.9\%) & 79 (53.0\%)&  58 (86.6\%) &  75 (68.2\%) \\
    
     \cmidrule(l){1-1} \cmidrule(l){2-2} \cmidrule(l){3-3}  \cmidrule(l){4-4}  \cmidrule(l){5-5}  \cmidrule(l){6-6}  \cmidrule(l){7-7} 
    
     \textbf{{Therapy purpose}} & \\
     
     {  }Definitive   & 270 (34.9\%) & 30 (17.8\%) & 82 ({59.9}\%) & 79 (53.0\%) & 58 (86.6\%) & 73 (66.4\%)\\
     {  }Postoperative   & 74 (9.6\%) & 19 (11.2\%) & 14 (10.2\%) & 3 (2.0\%) & 3 (4.5\%) & 24 (21.8\%)  \\
     {  }Salvage   & 431 (55.7\%) & 120 (71.0\%) & 41 ({29.9}\%) & 67 (45.0\%) & 6 (9.0\%) &  13 (11.8\%)  \\
    
     \cmidrule(l){1-1} \cmidrule(l){2-2} \cmidrule(l){3-3}  \cmidrule(l){4-4}  \cmidrule(l){5-5}  \cmidrule(l){6-6}  \cmidrule(l){7-7} \cmidrule(l){7-7} 

     \textbf{{CT Scanner}} &       &       &  \\ 
     {  }Manufacturer & Canon & Canon  &  Canon &  {SIEMENS} & GE&  {SIEMENS} \\
     {  }Model & Aquilion LB & Aquilion LB  & Aquilion LB &  {SOMATOM} & Discovery RT & {SOMATOM} \\
     {  }Scan mode & Helical & Helical & Helical & Helical & Helical & Helical\\
     {  }Filter type & LARGE & LARGE & LARGE & FLAT & BODY & FLAT \\
     {  }kVp & 120 & 120 & 120 & 120 & 140 & 120 \\
     {  }Spatial pixel size (mm) & 0.977 & 0.977 & 1.367 & 1.269 & 0.977& 1.269  \\
     {  }Slice thickness (mm) &  2 & 2 & 3 & 5 & 1.25 & 2  \\ 
    
     \bottomrule
      \multicolumn{7}{l}{{{Note.} {$^{\dagger}$ indicates utilized samples for each 1-shot / 2-shots / 3-shots training for each prostate specific antigen (PSA) cluster. PSA clusters (0-4) are categorized as:}}}\\
      \multicolumn{7}{l}{{0 ($\mathrm{PSA} < 5.0$), 1 ($5.0 \leq \mathrm{PSA} < 10.0$), 2 ($10.0 \leq \mathrm{PSA} < 20.0$), 3 ($20.0 \leq \mathrm{PSA} < 30.0$), and 4 ($\mathrm{PSA} \geq 30.0$)}}
     \end{tabular}
     }
   \label{tab_dataset_prostate}
 \end{table*}%

 \clearpage


 \begin{table*}[!t]
   \centering
   \caption{{Examples of the curated prostate cancer clinical data from electronic medical records (EMR) data.}}
   \resizebox{0.8\linewidth}{!}{
     \begin{tabular}{lll}

     \toprule

      \bf{Center} & \multicolumn{1}{l}{\bf{EMR Data \blue{(Parsed Information)}}} & \multirow{1}{*}{\shortstack[c]{{\bf{Input Clinical Data}}}}  \\
     \toprule
    
      {A,B,C} & \multirow{1}{*}{\shortstack[l]{
      {61-years old patient.}\\
      {\#1. Prostate, Adenoca, \blue{GS7(4+3)}, \blue{pT3aN0}M0, Stage IIIB}\\
      {- Tumor location: Both lobes [Index tumor: right, posterior, volume (1.44cc)]}\\
      {- Extraprostatic extension: Present, focal (right posterior, width: 3.0mm, depth: 0.5mm)}\\ 
      {- Intraglandular tumor volume: V2 (2.64cc)}\\
      {- Lymphovascular invasion: Not identified}\\ 
      {- Prostatic intraepithelial neoplasia, high grade: Present}\\
      {}\\
      {...} \\
      {} \\
      {- Vas deferens, right: Free of ca}\\
      {- Vas deferens, left: Free of ca Seminal vesicle, right: Free of ca Seminal vesicle}\\
      {- \blue{LN (-)}, Bone (-) ** \blue{iPSA : 8.31} }\\
      {** Roach score : ECE 52.47  SV 18.31  LN 15.54}\\
      {s/p Prostate biopsy}\\
      {s/p RALRP } \\
      {\#2. {Recurrence}, prostate PSA elevation}\\
      {@ Prostate MRI No evidence of local recurrence No enlarged LNs on both iliac chain}\\
      {@ \blue{PSA 0.72} - 0.43 - 0.08 - 0.01} 
      }}
      & \multirow{1}{*}{\shortstack[l]{\\{$<$Grade$>$ 7 (4+3)}\\{$<$Stage$>$ pT3a, N0} \\{$<$Metastasis$>$ negative} \\ {$<$Age$>$ 61} \\{$<$PSA$>$ 8.31}}} \\ \\ \\ \\ \\ \\ \\ \\ \\ \\ \\ \\ \\ \\ \\ \\  \\ \\ \\ 

     \cmidrule(l){1-1} \cmidrule(l){2-2}  \cmidrule(l){3-3} 
     {D} &  \multirow{1}{*}{\shortstack[l]{{Tumor markers: \blue{Gleason 4+3 = 7}}\\{Clinical staging: \blue{cT1c} \blue{N0} M0 11.63 IIC}\\{Notes: \blue{69 y.o.} male with HTN/HLD, orthostatic hypotension, currently on Midodrine} \\ {...}\\{\blue{PSA 11.63} prostate cancer, with MRI showing a 73 cc prostate and stable 13 mm index } \\ {area (PIRADS 3 previously) in the right anterior transition zone at apex and PET CT}}} 
     & \multirow{1}{*}{\shortstack[l]{\\{$<$Grade$>$ 7 (4+3)}\\{$<$Stage$>$ cT1c, N0} \\{$<$Metastasis$>$ unknown} \\ {$<$Age$>$ 69} \\{$<$PSA$>$ 11.63}}} \\  \\ \\ \\ \\ \\  

     \cmidrule(l){1-1} \cmidrule(l){2-2}  \cmidrule(l){3-3} 
     {E} &  \multirow{1}{*}{\shortstack[l]{{diagnosis details: \blue{78}-year-old male with a history of rectal cancer status post neoadjuvant}\\{chemoradiation.}\\{: \blue{Gleason 5+4} prostate cancer, \blue{PSA 38.4}, \blue{cT3aN0}M0 }\\ {(rectal stenosis unable to do DRE but no T3 per MRI). }\\ {...}\\{Plan PBT 79.2Gy/44fx +18 mo ADT}}}
     & \multirow{1}{*}{\shortstack[l]{\\{$<$Grade$>$ 9 (5+4)}\\{$<$Stage$>$ cT3a, N0} \\{$<$Metastasis$>$ unknown} \\ {$<$Age$>$ 78} \\{$<$PSA$>$ 38.4}}} \\  \\ \\  \\ \\ \\  

     \bottomrule

     \end{tabular}%
     }
   \label{tab_inputtext}%
 \end{table*}%



\begin{table}[!h]
\caption{Sensitivity of $Diff_{(MoME)}$ to the per-center sample size.}
\centering
\small
\resizebox{0.7\linewidth}{!}{\begin{tabular}{lccccccc}
\toprule

\multirow{2}{*}{\bf{Test Center}} & \multirow{2}{*}{\bf{Center-specific Inference ($c$)}} & \multicolumn{5}{c}{\bf{Sample size ($N$)}} & \multirow{2}{*}{\bf{Pseudo Center ($\hat{c}$)}}\\
\cmidrule(l){3-7} 
 & & 5 & 10 & 15 & 20 & 75 \\

\midrule

\multirow{3}{*}{\shortstack[l]{\bf{Center A}\\\bf{(Validation set)}}}
& Center A & \bf{0.099} & \bf{0.025} & \bf{0.015} & \bf{0.015} & \bf{0.014} & \checkmark\\
& Center B & 0.707 & 0.633 & 0.628 & 0.631 & 0.625 \\
& Center C & 0.666 & 0.605 & 0.600 & 0.600 & 0.608 \\

\cmidrule(l){1-1} \cmidrule(l){2-2} \cmidrule(l){3-7} \cmidrule(l){8-8}

\multirow{3}{*}{\bf{Center D}}
& Center A & 0.861 & 0.844 & 0.851 & 0.854 & -- \\
& Center B & 0.525 & 0.532 & 0.523 & 0.530 & -- \\
& Center C & \bf{0.439} & \bf{0.461} & \bf{0.427} & \bf{0.439} & -- & \checkmark \\

\cmidrule(l){1-1} \cmidrule(l){2-2} \cmidrule(l){3-7} \cmidrule(l){8-8}

\multirow{3}{*}{\bf{Center E}}
& Center A & 0.773 & 0.777 & 0.776 & 0.774 & -- \\
& Center B & 0.519 & 0.533 & 0.550 & 0.563 & -- \\
& Center C & \bf{0.497} & \bf{0.505} & \bf{0.512} & \bf{0.521} & -- & \checkmark\\

\bottomrule
\multicolumn{8}{l}{Note. \textbf{Bold} indicates the minimum $Diff_{(MoME)}$ per test center (lower is better), identifying the assigned pseudo center ($\hat{c}$).} \\
\end{tabular}
}
\label{table:sensitivity}
\end{table}

\begin{table*}[!h]
\caption{PTV delineation performance comparison on MoME without fine-tuning (0-shot) and comparison between few-shot fine-tuning with external-center pretraining (Ours) vs.\ training from scratch on the closed-center data only (Baseline).}
\centering
\resizebox{0.85\linewidth}{!}{
\begin{tabular}{llcccccccc}
\toprule

\multirow{2}{*}{\bf{Closed Center}}
& \multirow{2}{*}{\bf{Metric}}
& \multirow{2}{*}{\bf{0-shot}}
& \multicolumn{2}{c}{\bf{1-shot}}
& \multicolumn{2}{c}{\bf{2-shots}}
& \multicolumn{2}{c}{\bf{3-shots}} \\
\cmidrule(l){4-5} \cmidrule(l){6-7} \cmidrule(l){8-9}
& & & {Ours} & {Baseline} & {Ours} & {Baseline} & {Ours} & {Baseline} \\

\midrule

\multirow{3}{*}{\shortstack[l]{\bf{Center D}\\\bf{(n=47)}}}
& {Dice $\uparrow$}
& \shortstack[c]{\bf{0.605}\\(0.577-0.629)}
& \shortstack[c]{\bf{0.628}\\(0.581-0.673)} & \shortstack[c]{0.310\\(0.232-0.399)}
& \shortstack[c]{\bf{0.682}\\(0.642-0.722)} & \shortstack[c]{0.474\\(0.417-0.536)}
& \shortstack[c]{\bf{0.677}\\(0.637-0.716)} & \shortstack[c]{0.553\\(0.494-0.611)} \\

& {IoU $\uparrow$}
& \shortstack[c]{\bf{0.439}\\(0.412-0.464)}
& \shortstack[c]{\bf{0.476}\\(0.430-0.520)} & \shortstack[c]{0.217\\(0.154-0.286)}
& \shortstack[c]{\bf{0.533}\\(0.491-0.577)} & \shortstack[c]{0.335\\(0.283-0.392)}
& \shortstack[c]{\bf{0.526}\\(0.485-0.568)} & \shortstack[c]{0.407\\(0.355-0.460)} \\

& {HD-95 $\downarrow$}
& \shortstack[c]{\bf{6.369}\\(5.002-7.792)}
& \shortstack[c]{\bf{6.606}\\(5.285-7.995)} & \shortstack[c]{27.574\\(26.934-28.185)}
& \shortstack[c]{\bf{5.161}\\(3.782-6.641)} & \shortstack[c]{25.617\\(24.812-26.415)}
& \shortstack[c]{\bf{4.283}\\(3.413-5.163)} & \shortstack[c]{20.333\\(19.564-21.133)} \\

\midrule

\multirow{3}{*}{\shortstack[l]{\bf{Center E}\\\bf{(n=90)}}}
& {Dice $\uparrow$}
& \shortstack[c]{\bf{0.490}\\(0.460-0.522)}
& \shortstack[c]{\bf{0.568}\\(0.534-0.600)} & \shortstack[c]{0.255\\(0.228-0.284)}
& \shortstack[c]{\bf{0.596}\\(0.563-0.626)} & \shortstack[c]{0.364\\(0.328-0.399)}
& \shortstack[c]{\bf{0.612}\\(0.573-0.646)} & \shortstack[c]{0.489\\(0.448-0.523)} \\

& {IoU $\uparrow$}
& \shortstack[c]{\bf{0.338}\\(0.312-0.365)}
& \shortstack[c]{\bf{0.413}\\(0.382-0.444)} & \shortstack[c]{0.154\\(0.134-0.175)}
& \shortstack[c]{\bf{0.441}\\(0.409-0.469)} & \shortstack[c]{0.235\\(0.208-0.262)}
& \shortstack[c]{\bf{0.461}\\(0.425-0.494)} & \shortstack[c]{0.343\\(0.306-0.373)} \\

& {HD-95 $\downarrow$}
& \shortstack[c]{\bf{2.810}\\(2.267-3.417)}
& \shortstack[c]{\bf{2.049}\\(1.635-2.550)} & \shortstack[c]{23.525\\(22.837-24.189)}
& \shortstack[c]{\bf{1.766}\\(1.396-2.248)} & \shortstack[c]{22.258\\(20.952-23.443)}
& \shortstack[c]{\bf{1.788}\\(1.400-2.299)} & \shortstack[c]{3.094\\(2.297-4.044)} \\

\bottomrule
\multicolumn{9}{l}{Note. \textbf{Bold} indicates best performance per shot setting.}
\end{tabular}
}
\label{tab_finetune_analysis}
\end{table*}

 \begin{table*}[!t]
   \centering
   \caption{{Details of SegRap2025 nasopharyngeal cancer data partitioning and characteristics for each cohort.}}
   \resizebox{0.95\linewidth}{!}{
     \begin{tabular}{lcccccc}
     \toprule
    
     \multirow{1}{*}{\shortstack[c]{\bf{Center}}} & \multicolumn{2}{c}{\textbf{Internal Cohort}} & \multicolumn{1}{c}{\textbf{Cohort \#1}} & \multicolumn{1}{c}{{\bf{Cohort \#2}}} & \multicolumn{1}{c}{{\bf{Closed Cohort \#3}}} & \multicolumn{1}{c}{{\bf{Closed Cohort \#4}}}  \\
      \cmidrule(l){1-1} \cmidrule(l){2-3}  \cmidrule(l){4-4}  \cmidrule(l){5-5}  \cmidrule(l){6-6}  \cmidrule(l){7-7} 
     
      \shortstack[c]{\textbf{{Hospital}}\\{}} & \multicolumn{2}{c}{\shortstack[c]{{Sichuan Cancer Hospital}\\{}}}  & \shortstack[c]{{Sichuan Provincial}\\{People's Hospital}} & \shortstack[c]{{The First Affiliated}\\{Hospital of USTC}} & \shortstack[c]{{Southern Medical}\\{University}} & \shortstack[c]{{Daguan Hospital of}\\{Chengdu Jinjiang}} \\
      \cmidrule(l){1-1} \cmidrule(l){2-3}  \cmidrule(l){4-4}  \cmidrule(l){5-5}  \cmidrule(l){6-6}  \cmidrule(l){7-7} 
     
      \multirow{2}{*}{\shortstack[c]{\bf{Data split}}} & \multirow{2}{*}{{Train (n=240)}} 
      & \multirow{2}{*}{{Test (n=60)}}
      & \multicolumn{1}{c}{{Train (n=2/3/4$^{\dagger}$)}} 
      & \multicolumn{1}{c}{{{Train (n=2/3/4$^{\dagger}$)}}} 
      \\

      &
      &  
      & \multicolumn{1}{c}{{Test (n=56)}} 
      & \multicolumn{1}{c}{{{Test (n=29)}}} 
      & \multicolumn{1}{c}{{{Test (n=20)}}}
      & \multicolumn{1}{c}{{{Test (n=20)}}} \\
    
     \cmidrule(l){1-1} \cmidrule(l){2-2} \cmidrule(l){3-3}  \cmidrule(l){4-4}  \cmidrule(l){5-5}  \cmidrule(l){6-6}  \cmidrule(l){7-7} 

     \shortstack[c]{\textbf{{Label Description}}\\{}} & \\
     {  }0: Background   &\\
     {  }1: $^{\ddagger}L_{Ib}$   &\\
     {  }2: $^{\ddagger}L_{II+III+Va}$   &\\
     {  }3: $^{\ddagger}L_{IV+Vb+Vc}$   &\\
     {  }4: $^{\ddagger}R_{Ib}$   &\\
     {  }5: $^{\ddagger}R_{II+III+Va}$   &\\
     {  }6: $^{\ddagger}R_{IV+Vb+Vc}$   &\\
     \cmidrule(l){1-1} \cmidrule(l){2-2} \cmidrule(l){3-3}  \cmidrule(l){4-4}  \cmidrule(l){5-5}  \cmidrule(l){6-6}  \cmidrule(l){7-7}

     \textbf{{CT Scanner}} &       &       &  \\ 
     {  }Manufacturer & Philips  & {Philips} & {SIEMENS}  &  {SIEMENS} &  {SIEMENS} & {SIEMENS} \\
     {  }Model & Brilliance Big Bore & {Brilliance Big Bore}  & {SOMATOM} &  {SOMATOM} & {SOMATOM} & {SOMATOM} \\
     {  }kVp & 120 & 120 & 120-140 & 120-140 & 120-140 & 120 \\
     {  }Current (mA) & 275-375 & 275-375 & 280-380 & 280-380 & 280-380 & 200-250   \\
     {  }Slice thickness (mm) &  3 & 3 & 3 & 3 & 3 & 2.5  \\ 
    
     \bottomrule
     \multicolumn{7}{l}{{{Note.} {$^{\dagger}$ indicates utilized samples for each 1-shot / 2-shots / 3-shots training with 1-shot validation, $^{\ddagger}$ indicates lymph node (LN) labels; $L_{Ib}$: Left level Ib}}} \\
     \multicolumn{7}{l}{{LNs, $L_{II+III+Va}$: Left levels II, III, and Va LNs, $L_{IV+Vb+Vc}$: Left levels IV, Vb, and Vc LNs, $R_{Ib}$: Right level Ib LNs, $R_{II+III+Va}$: Right levels II, }} \\
     \multicolumn{7}{l}{{III, and Va LNs, $R_{IV+Vb+Vc}$: Right levels IV, Vb, and Vc LNs}} \\

     \end{tabular}
     }
   \label{tab_dataset_prostate}
 \end{table*}%

 \begin{table}[!t]
 \centering
 \caption{{Average CTV delineation performance across 6 labels in nasopharyngeal cancer.}}
 \resizebox{0.7\linewidth}{!}{
     \begin{tabular}{llccc}
     \toprule
    
     \multirow{3}{*}{\bf{Dataset}} 
     & \multirow{3}{*}{\bf{Metric}} 
     & \multicolumn{1}{c}{\bf{(a) Single-center}} & \multicolumn{2}{c}{\bf{(b) Multicenter AI Training}} \\
    
     \cmidrule(l){3-3} \cmidrule(l){4-5}  
     & & \multirow{1}{*}{{3D ResUNet\cite{cciccek20163d}}} &   
     \multicolumn{1}{c}{{3D ResUNet \cite{cciccek20163d}}} & \multicolumn{1}{c}{{MoME (Ours)}} \\
    
     \midrule  
     \multirow{4}{*}{\shortstack[l]{\bf{Internal Cohort}\\\bf{(n=60)}}}  & \shortstack[c]{{Dice $\uparrow$}\\{ }}   & \shortstack[c]{{0.607}\\{(0.600-0.615)}} & \shortstack[c]{{0.730}\\{(0.720-0.738)}} & \shortstack[c]{\bf{0.742}\\{(0.733-0.752)}}  \\ 
     & \shortstack[c]{{IoU $\uparrow$}\\{ }}   & \shortstack[c]{{0.493}\\{(0.485-0.501)}} & \shortstack[c]{{0.590}\\{(0.580-0.600)}} & \shortstack[c]{\bf{0.604}\\{(0.594-0.615)}}  \\ 
     & \shortstack[c]{{HD-95 $\downarrow$}\\{ }}  & \shortstack[c]{{2.316}\\{(2.219-2.406)}} & \shortstack[c]{{0.329}\\{(0.299-0.363)}} & \shortstack[c]{\bf{0.323}\\{(0.292-0.360)}}  \\ 
    
     \cmidrule(l){1-1} \cmidrule(l){2-2} \cmidrule(l){3-3} \cmidrule(l){4-4}  \cmidrule(l){5-5}  \multirow{4}{*}{\shortstack[l]{\bf{Cohort \#1}\\\bf{(n=56)}}} & \shortstack[c]{{Dice $\uparrow$}\\{ }}   & \shortstack[c]{{0.586}\\{(0.578-0.593)}} & \shortstack[c]{{0.694}\\{(0.686-0.702)}} & \shortstack[c]{\bf{0.716}\\{(0.708-0.723)}}  \\ 
     & \shortstack[c]{{IoU $\uparrow$}\\{ }}   & \shortstack[c]{{0.467}\\{(0.459-0.474)}} & \shortstack[c]{{0.546}\\{(0.537-0.554)}} & \shortstack[c]{\bf{0.569}\\{(0.559-0.578)}}  \\ 
     & \shortstack[c]{{HD-95 $\downarrow$}\\{ }}  & \shortstack[c]{{2.562}\\{(2.412-2.743)}} & \shortstack[c]{{3.724}\\{(2.814-4.677)}} & \shortstack[c]{\bf{2.171}\\{(1.269-3.214)}}  \\ 
    
     \cmidrule(l){1-1} \cmidrule(l){2-2} \cmidrule(l){3-3} \cmidrule(l){4-4}  \cmidrule(l){5-5}  \multirow{4}{*}{\shortstack[l]{\bf{Cohort \#2}\\\bf{(n={29})}}} & \shortstack[c]{{Dice $\uparrow$}\\{ }}   & \shortstack[c]{{0.596}\\{(0.587-0.604)}} & \shortstack[c]{{0.692}\\{(0.680-0.704)}} & \shortstack[c]{\bf{0.700}\\{(0.690-0.710)}}  \\ 
     & \shortstack[c]{{IoU $\uparrow$}\\{ }}   & \shortstack[c]{{0.475}\\{(0.465-0.485)}} & \shortstack[c]{{0.545}\\{(0.532-0.559)}} & \shortstack[c]{\bf{0.551}\\{(0.539-0.563)}}  \\ 
     & \shortstack[c]{{HD-95 $\downarrow$}\\{ }}  & \shortstack[c]{\bf{3.851}\\{(3.136-4.615)}} & \shortstack[c]{{4.327}\\{(2.261-6.635)}} & \shortstack[c]{{7.207}\\{(3.459-11.146)}}  \\ 
    
     \cmidrule(l){1-1} \cmidrule(l){2-2} \cmidrule(l){3-3} \cmidrule(l){4-4}  \cmidrule(l){5-5}  \multirow{4}{*}{\shortstack[l]{\bf{Closed Cohort \#3}\\\bf{(n={20})}}} & \shortstack[c]{{Dice $\uparrow$}\\{ }}   & \shortstack[c]{{0.584}\\{(0.570-0.596)}} & \shortstack[c]{{0.708}\\{(0.693-0.721)}} & \shortstack[c]{\bf{0.723}\\{(0.707-0.738)}}  \\ 
     & \shortstack[c]{{IoU $\uparrow$}\\{ }}   & \shortstack[c]{{0.470}\\{(0.455-0.483)}} & \shortstack[c]{{0.563}\\{(0.546-0.578)}} & \shortstack[c]{\bf{0.581}\\{(0.562-0.598)}}  \\ 
     & \shortstack[c]{{HD-95 $\downarrow$}\\{ }}  & \shortstack[c]{{2.391}\\{(1.955-2.918)}} & \shortstack[c]{\bf{0.457}\\{(0.339-0.662)}} & \shortstack[c]{{0.863}\\{(0.346-1.631)}}  \\ 
    
     \cmidrule(l){1-1} \cmidrule(l){2-2} \cmidrule(l){3-3} \cmidrule(l){4-4}  \cmidrule(l){5-5}  \multirow{4}{*}{\shortstack[l]{\bf{Closed Cohort \#4}\\\bf{(n={20})}}} & \shortstack[c]{{Dice $\uparrow$}\\{ }}   & \shortstack[c]{{0.564}\\{(0.547-0.583)}} & \shortstack[c]{{0.695}\\{(0.675-0.715)}} & \shortstack[c]{\bf{0.712}\\{(0.693-0.733)}}  \\ 
     & \shortstack[c]{{IoU $\uparrow$}\\{ }}   & \shortstack[c]{{0.448}\\{(0.427-0.469)}} & \shortstack[c]{{0.546}\\{(0.522-0.569)}} & \shortstack[c]{\bf{0.565}\\{(0.540-0.590)}}  \\ 
     & \shortstack[c]{{HD-95 $\downarrow$}\\{ }}  & \shortstack[c]{{2.573}\\{(2.161-3.107)}} & \shortstack[c]{{0.645}\\{(0.414-0.991)}} & \shortstack[c]{\bf{0.553}\\{(0.378-0.801)}} \\ 
         
     \bottomrule
     \multicolumn{5}{l}{{{Note.} {\textbf{Bold} metric indicates best performance. All experimental results are from  3-shot setting.}}}
    
     \end{tabular}
 }
 \label{tab_naso}
 \end{table}

\begin{table}[!h]
\caption{PTV delineation performance: one-step (Ours) vs.\ step-wise (Prostate $\rightarrow$ PTV) in Dice Performance.}
\centering
\small
\begin{tabular}{llcccc}
\toprule
\multirow{2}{*}{\bf{Method}}
& \multirow{2}{*}{\bf{Prostate Segmmentation Module}}
& \multirow{2}{*}{\bf{Center A}}
& \multirow{2}{*}{\bf{Center C}}
& \multicolumn{2}{c}{\bf{Closed Center D}} \\
\cmidrule(l){5-6}
 & & & & 0-shot & 3-shot \\

\midrule

\shortstack[c]{\bf{One-step PTV (Ours)}\\{ }} & \shortstack[c]{--\\{ }} 
& \shortstack[c]{\bf{0.756}\\(0.731-0.778)}
& \shortstack[c]{0.692\\(0.661-0.725)}
& \shortstack[c]{\bf{0.605}\\(0.577-0.629)}
& \shortstack[c]{\bf{0.677}\\(0.637-0.716)} \\

\midrule

\multirow{2}{*}{\bf{\shortstack[l]{Step-wise\\(Prostate $\rightarrow$ PTV)}}}
& \shortstack[c]{3D U-Net (in-house dataset)\\{ }} 
& \shortstack[c]{0.747\\(0.721-0.772)}
& \shortstack[c]{\bf{0.716}\\(0.688-0.744)}
& \shortstack[c]{0.465\\(0.419-0.516)}
& \shortstack[c]{0.582\\(0.529-0.633)} \\

& \shortstack[c]{Public TotalSegmentator$^\dagger$\\{ }}  
& \shortstack[c]{\bf{0.765}\\(0.741-0.788)}
& \shortstack[c]{0.672\\(0.640-0.704)}
& \shortstack[c]{0.551\\(0.503-0.598)}
& \shortstack[c]{0.604\\(0.557-0.651)} \\

\bottomrule
\multicolumn{6}{l}{$^\dagger$\footnotesize https://github.com/wasserth/totalsegmentator}
\end{tabular}
\label{tab_onestep_vs_stepwise}
\end{table}

 \begin{figure}[!h]
 \centering
 \includegraphics[width=1\linewidth]{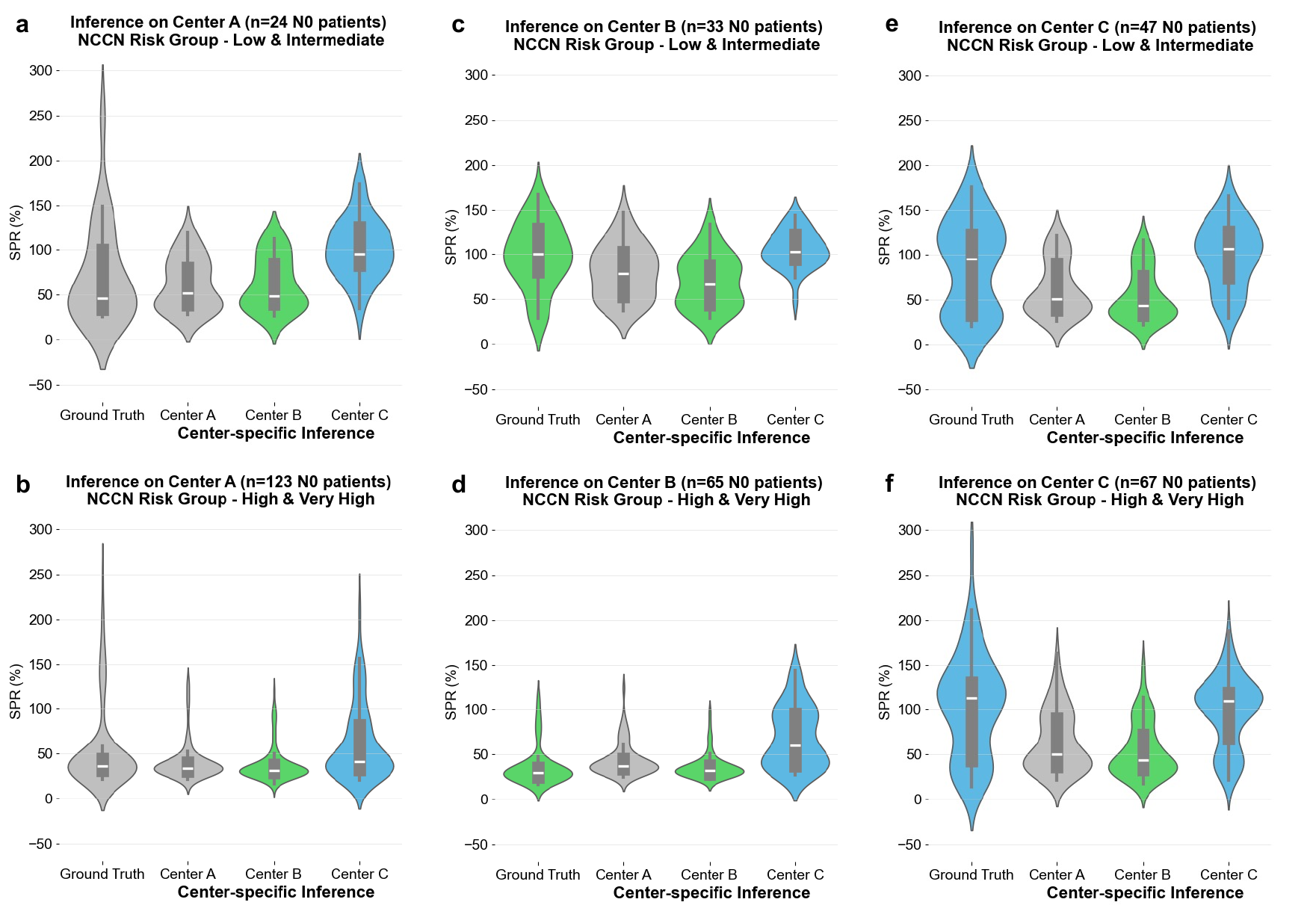}%
 \caption{Comparison of Sacrum-to-PTV ratio (SPR) across risk groups when using each center-specific router. For Center A, both (a) low/intermediate-risk and (b) high/very high-risk groups show the closest alignment with the ground truth SPR distribution when using the Center A expert router, while the Center B expert router produces similar results. In contrast, using the Center C expert router leads to the largest deviation, effectively highlighting the similarities and differences in practice patterns across institutions. For Center B, the (c) low/intermediate-risk and (d) high/very high-risk groups also show consistent alignment with Center B’s original practice when using the Center B expert router, with a similarly close match from the Center A expert router, whereas the Center C expert router again leads to a significant increase in SPR. In Center C, both (e) low/intermediate-risk and (f) high/very high-risk groups display a distinct pattern, with higher SPR values that reflect the center's less frequent use of PNI and tighter PTV margins.}
 \label{fig_supp_risk_experts}
 \end{figure}

 \begin{figure*}[!h]
 \centering
 \includegraphics[width=1\linewidth]{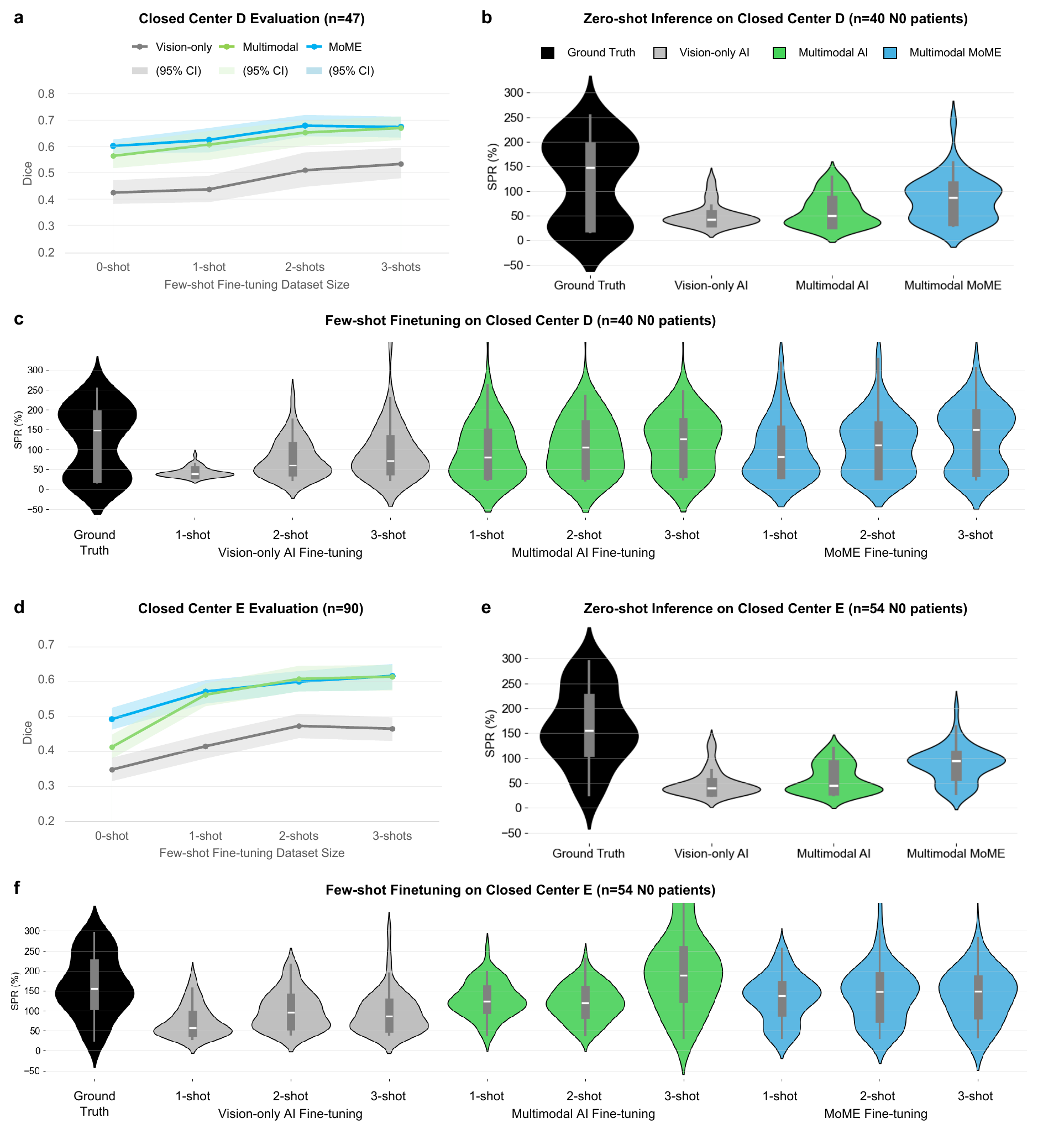}%
 \caption{{Closed center dataset evaluation. (a) Closed Center D evaluation in Dice metric with varying few-shot fine-tuning dataset sizes. (b) SPR distribution of zero-shot inference for each method, and (c) few-shot fine-tuning result with varying number of few-shot fine-tuning of the closed center D dataset. For (a-b), the Dice metric for each trial is presented as mean values (center lines) with 95th percentile of confidence intervals (shaded areas). (d) Closed Center E evaluation in Dice metric based on varying few-shot fine-tuning dataset sizes. (e) SPR distribution of zero-shot inference for each method, and (c) few-shot fine-tuning result with varying number of few-shot fine-tuning of the closed center E dataset.}} %
 \label{fig_finetune}
 \end{figure*}

 \begin{figure}[!h]
 \centering
 \includegraphics[width=1\linewidth]{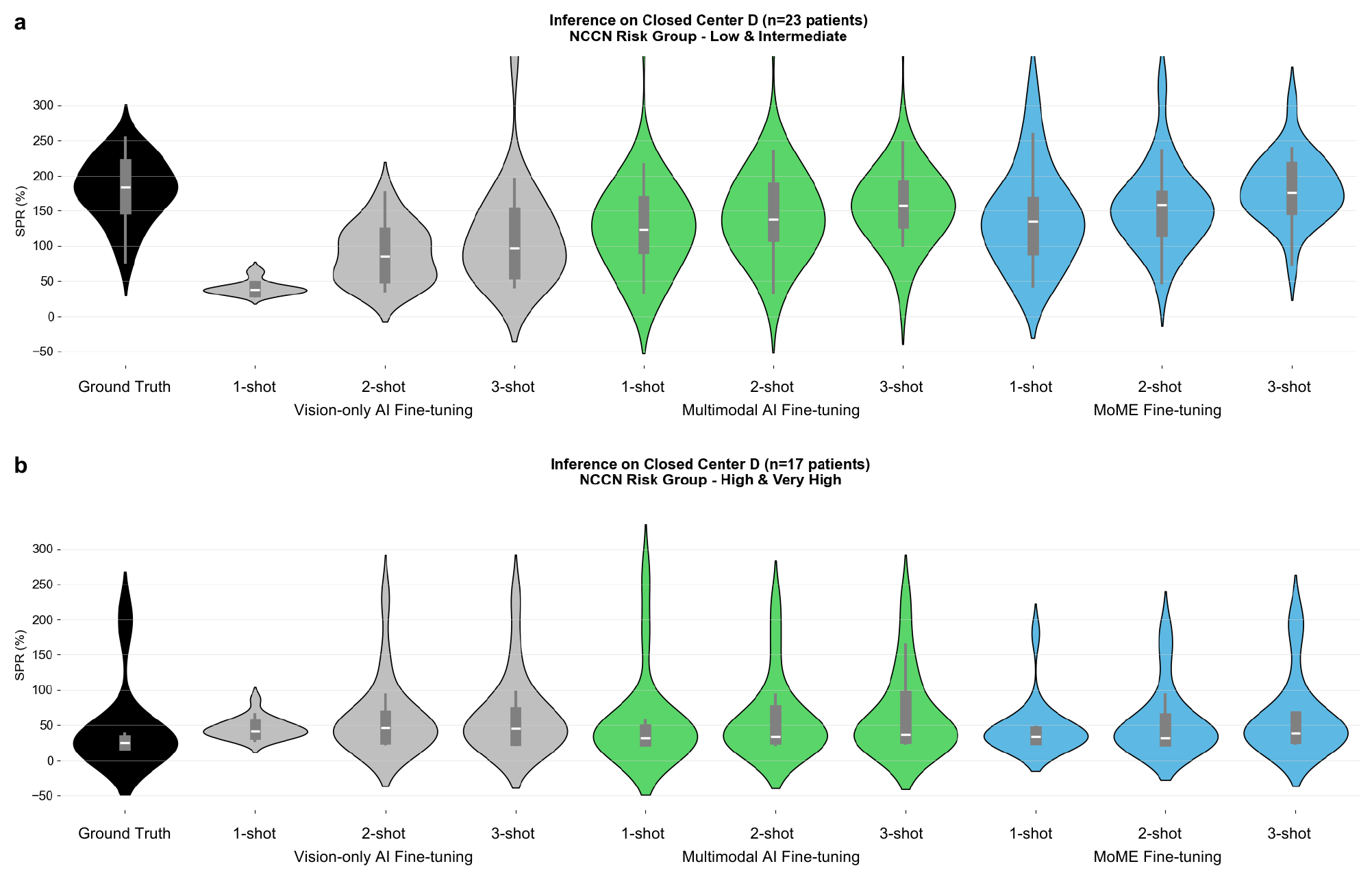}%
 \caption{{Comparison of Sacrum-to-PTV ratio (SPR) across risk groups for the closed center D. In both (a) low/intermediate-risk and (b) high/very high-risk groups, the MoME model demonstrates a closer alignment with the ground truth distribution compared to the multi-modal as well as the vision only models. This trend becomes more pronounced as the number of examples increases with 1-shot, 2-shot, and 3-shot learning. Notably, in the high-risk group, the SPR distribution produced by the MoME model nearly matches the ground truth with just three-shot fine-tuning. }}
 \label{fig_supp_risk_closed}
 \end{figure}

 \begin{figure}[!h]
 \centering
 \includegraphics[width=1\linewidth]{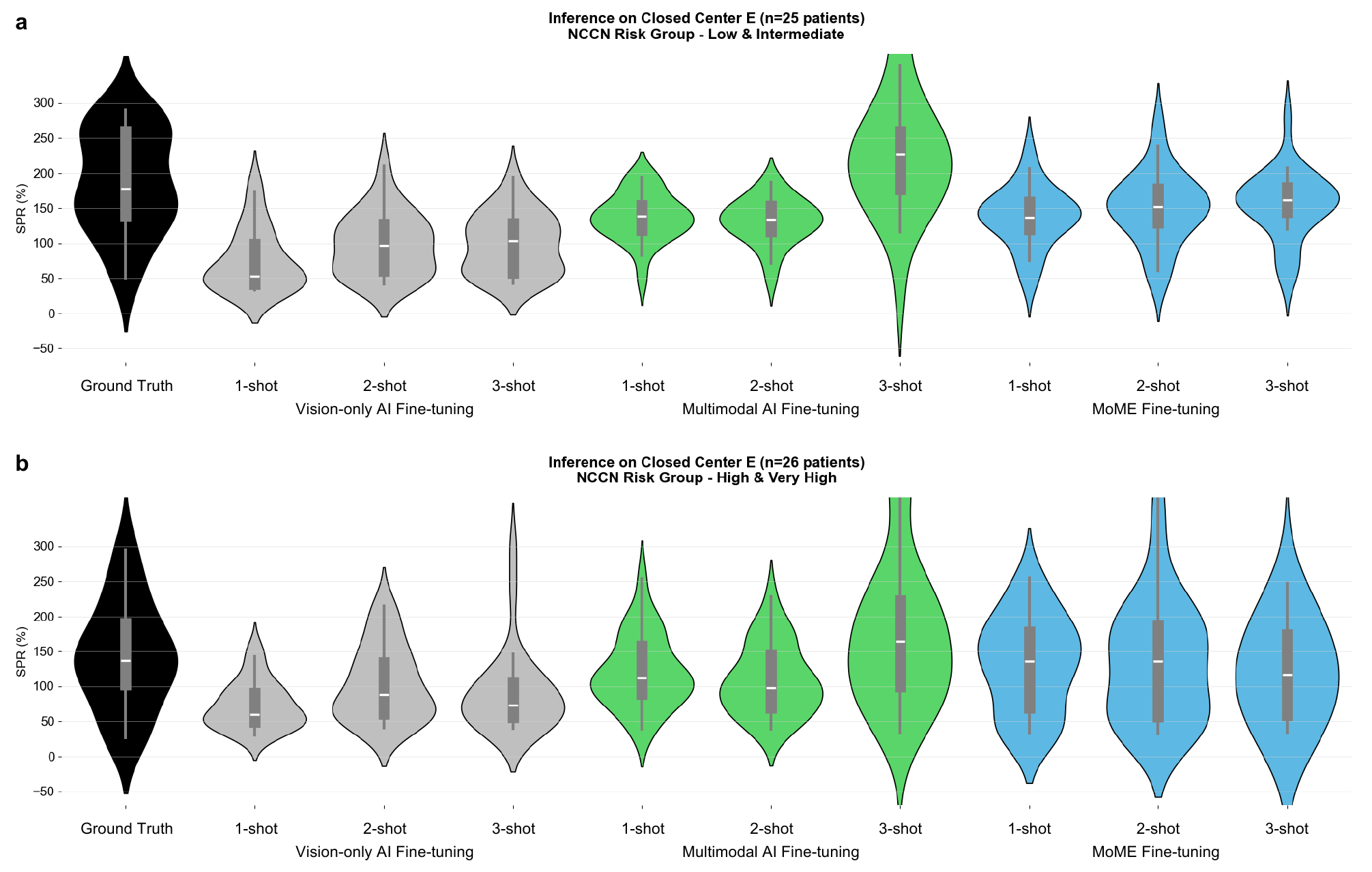}%
 \caption{{Comparison of Sacrum-to-PTV ratio (SPR) across risk groups for the closed center E. (a) low/intermediate-risk and (b) high/very high-risk groups, the MoME model demonstrates  the most similar distribution with the ground truth distribution compared to the multi-modal as well as the vision only models. The distribution gets more similar to the ground truth as the few-shot tuning samples get increased to 3-shot learning, specifically in the low \& intermediate group. }}
 \label{fig_supp_risk_closed_mayo}
 \end{figure}

 \begin{figure*}[!h]
 \centering
 \includegraphics[width=0.9\linewidth]{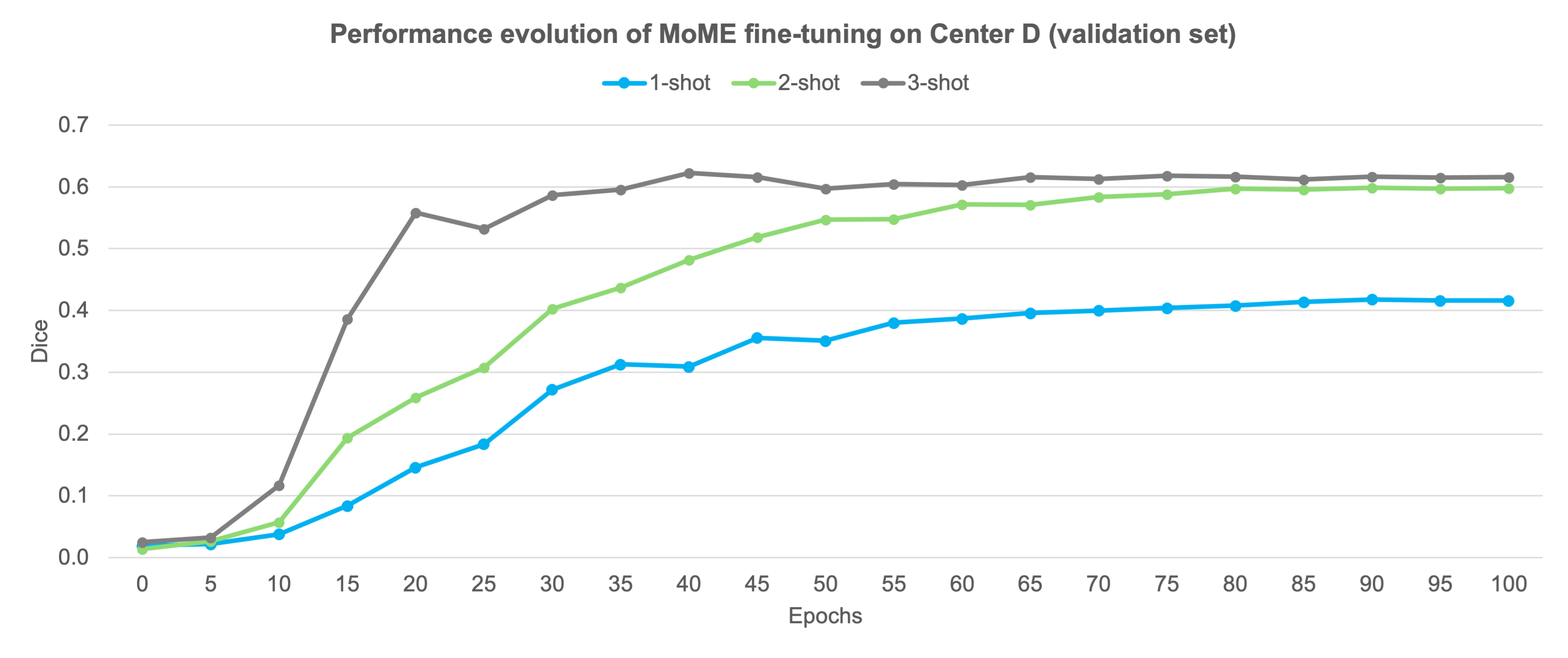}%
 \caption{{Performance evolution across fine-tuning epochs on Center D validation set. Patterns are consistent: (i) the Dice improves monotonically and saturates for every shot regime; (ii) convergence is faster with more shots, with the 3-shot setting reaching its plateau by about epoch 30 to 40, the 2-shot setting by about epoch 70 to 80, and the 1-shot setting by about epoch 90; and (iii) the asymptotic validation Dice scales with shot count. Beyond the saturation point, the curves flatten, so additional epochs add little, indicating that MoME fine-tunes efficiently with a small number of epochs.}} %
 \label{fig_finetune_epoch}
 \end{figure*}

\balance
\end{document}